\documentclass[prd,aps,preprint,showpacs,preprintnumbers,amsmath,amssymb,tighten,nofootinbib,12pt]{revtex4}

\usepackage{graphicx}
\usepackage{dcolumn}
\usepackage{bm}
\usepackage{bbm}

\begin{document}


\title{Dirac Lepton Angle Matrix v.s. Majorana Lepton Angle Matrix and Their Renormalization Group Running Behaviours
\vspace{0.3cm}}

\author{\bf Shu Luo}
\email{luoshu@xmu.edu.cn}

\affiliation{Department of Physics and Institute of Theoretical Physics and Astrophysics, \\ Xiamen University, Xiamen, Fujian, 361005 China\vspace{0.9cm}}

%

\begin{abstract}
Enlightened by the idea of the $3\times 3$ CKM angle matrix proposed recently by Harrison {\it et al.}, we introduce the Dirac angle matrix $\Phi$ and the Majorana angle matrix $\Psi$ in the lepton sector for Dirac and Majorana neutrinos respectively. We show that in the presence of CP violation, the angle matrix $\Phi$ or $\Psi$ is entirely equivalent to the complex MNS matrix $V$ itself, but has the advantage of being real, phase rephasing invariant, directly associated to the leptonic unitarity triangles (UTs) and do not depend on any particular parametrization of $V$. In this paper, we further analyzed how the angle matrices evolve with the energy scale. The one-loop Renormalization Group Equations (RGEs) of $\Phi$, $\Psi$ and some other rephasing invariant parameters are derived and the numerical analysis is performed to compare between the case of Dirac and Majorana neutrinos. Different neutrino mass spectra are taken into account in our calculation. We find that apparently different from the case of Dirac neutrinos, for Majorana neutrinos the RG-evolutions of  $\Phi$, $\Psi$ and ${\cal J}$ strongly depend on the Majorana-type CP-violating parameters and are more sensitive to the sign of $\Delta m^{2}_{31}$. They may receive significant radiative corrections in the MSSM with large $\tan\beta$ if three neutrino masses are nearly degenerate.
\end{abstract}

\pacs{14.60.Pq, 13.10.+q, 25.30.Pt}

\maketitle

\section{Introduction}

Since 1998, a number of successful neutrino oscillation experiments have provided us with very convincing evidence that neutrinos are massive and lepton flavors are mixed \cite{Nakamura:2010zzi}. The flavor mixing among three neutrinos can be described by the Maki-Nakagawa-Sakata (MNS) matrix $V$. Now three mixing angles in $V$ and two squared neutrino mass differences have been approximately determined. But whether neutrinos are Dirac or Majorana particles remains an open question. The Majorana nature of massive neutrinos can be revealed by the investigation of processes in which the total lepton charge $L$ changes by two units $\Delta L = 2$. Neutrinoless double-beta ($0\nu\beta\beta$) decay experiments are considered as the most promising method in this catalog. If $0\nu\beta\beta$ decay is eventually observed, we shall make sure that neutrinos are Majorana particles. If there is no experimental signal for the $0\nu\beta\beta$ decay, however, we shall be unable to conclude that neutrinos are just Dirac particles \cite{0nubetabeta}. 

CP violation in the lepton sector is another open question. In the framework of three Dirac neutrinos, CP violation in the MNS matrix $V$ can be described by a single Dirac CP-violating phase which can be measured in the neutrino oscillation experiments. If three neutrinos are Majorana particles, two extra Majorana CP-violating phases are introduced in $V$. It is well known that the presence of Majorana phases introduces some novel features in leptonic CP violation, like the possibility of having CP violation in the case of two Majorana neutrinos \cite{Schechter:1980gr} as well as having CP breaking even in the limit of three exactly degenerate neutrinos \cite{degenerate}. Besides, these extra Majorana phases can affect significantly the rates of $0\nu\beta\beta$ decay \cite{neutrinoless double-beta} and some LFV decays \cite{LFV}, play an important role in the RG-evolutions of the neutrino masses and mixing matrix $V$ \cite{Majorana v.s. Dirac}, and be the source of CP violation in the leptogenesis \cite{leptogenesis}. Although the Majorana phases can not be directly measured in neutrino oscillation experiments \cite{Bilenky:1980cx}, the constrains on them can be drawn indirectly from the studies of above mentioned processes.

If the $3 \times 3$ MNS matrix $V$ is unitary, its nine elements satisfy the following normalization and orthogonality conditions:
\begin{equation}
\sum_\alpha V^{}_{\alpha i} V^{*}_{\alpha j} \; = \; \delta^{}_{ij} \; , ~~~~ \sum_i V^{}_{\alpha i} V^{*}_{\beta i} \; = \; \delta^{}_{\alpha\beta} \; ,
\end{equation}
where the Greek and Latin subscripts run over $(e, \mu, \tau)$ and $(1, 2, 3)$, respectively. The six orthogonality relations geometrically define six leptonic unitarity triangles (UTs) in the complex plane \cite{UTs}, as illustrated in Fig. 1. The six UTs have eighteen different sides and nine different inner angles, but their areas are all identical to $\cal J$/2 with $\cal J$ being the Jarlskog invariant of CP violation \cite{Jarlskog} defined by
\begin{equation}
{\rm Im} \left( V^{}_{\alpha i} V^{}_{\beta j} V^{*}_{\alpha j} V^{*}_{\beta i} \right) \; = \; {\cal J} \sum_\gamma \epsilon^{}_{\alpha\beta\gamma} \sum_k \epsilon^{}_{ijk} \; .
\end{equation}
The leptonic UTs provide us with a visible way of studying the CP violation in the lepton sector. In the presence of CP violation, all the physical parameters in $V$ can be drawn from these six UTs no matter if the neutrinos are Dirac or Majorana particles \cite{UTs, Majorana UTs}.

Enlightened by the idea of the $3\times 3$ CKM angle matrix proposed recently by Harrison {\it et al.} \cite{Harrison:2009bz}, we introduce the $3\times 3$ Dirac angle matrix $\Phi$ in the lepton sector for Dirac neutrinos. Moreover, we extrapolate this concept to account for Majorana neutrinos and propose the Majorana angle matrix $\Psi$. In section II, we introduce the concepts of $\Phi$ and $\Psi$. We show that in the presence of CP violation, the angle matrix $\Phi$ ($\Psi$) is entirely equivalent to the complex mixing matrix $V$ itself for Dirac (Majorana) neutrinos, but has the advantage of being real, phase rephasing invariant, directly associated to the leptonic UTs and do not depend on any particular parametrization of the MNS matrix. In Section III, we further analyze how the angle matrices and some other rephasing invariant parameters evolve with the energy scale. The one-loop Renormalization Group Equations (RGEs) of $|V|$, $\Phi$, $\Psi$ and the Jarlskog ${\cal J}$ are derived. Unlike the CKM angle matrix which is quite stable against the RG-evolution \cite{Luo:2009wa}, the leptonic angle matrix $\Phi$ and $\Psi$ may receive significant radiative corrections when evolve from the electroweak scale $\Lambda^{}_{\rm EW}$ to a superhigh energy scale. Section IV is devoted to a numerical analysis of the RGE running behaviours of $\Phi$ and $\Psi$, and to a careful comparison between the case of Dirac and Majorana neutrinos. Different neutrino mass spectra are taken into account in our calculation. A brief summary of the main results is given in section V.

\section{Dirac Angle Matrix v.s. Majorana Angle Matrix}

In the mass eigenstate basis, the charged-current interaction of leptons is described by
\begin{equation}
{\cal L}^{}_{\rm CC} = \frac{g}{\sqrt{2}} \; \overline{\left ( \begin{matrix} l^{}_{1} & l^{}_{2} & l^{}_{3} \end{matrix} \right )^{}_{\rm L}} \; \gamma^{\mu}_{} \; V \left ( \begin{matrix} \nu^{}_{1} \cr \nu^{}_{2} \cr \nu^{}_{3} \end{matrix} \right )^{}_{\rm L} W^{-}_{\mu} \; + \; {\rm h.c.}  \; ,
\end{equation}
and the CP violation is naturally included if the MNS matrix $V$ is complex. If the neutrinos are Dirac particles, one has the freedom to make phase rotations on both the charged lepton fields and the neutrino fields, which leads to the redefinition
\begin{equation}
V \rightarrow e^{i \Phi^{}_{l}} V e^{-i \Phi^{}_{\nu}} \; ,
\end{equation}
where $\Phi^{}_{l} = {\rm diag} (\phi^{}_{l^{}_{1}}, \phi^{}_{l^{}_{2}}, \phi^{}_{l^{}_{3}})$ and  $\Phi^{}_{\nu} = {\rm diag} (\phi^{}_{\nu^{}_{1}}, \phi^{}_{\nu^{}_{2}}, \phi^{}_{\nu^{}_{3}})$. Physical quantities are basis independent and must be invariant under the rephasing of Eq. (4). The simplest rephasing invariant quantities are the moduli of nine elements of $V$ which are all real. Although they might implicitly involve CP-violating phases, it is convenient to look for imaginary parameters that explicitly require CP violation. As been pointed by many references, the lowest-order (in  $V$) rephasing invariants that are not automatically real are the quartic products (which are also called ``boxes" in some references) \cite{Box}:
\begin{equation}
^{\alpha i}_{}\Box^{}_{\beta j} \; \equiv \; V^{}_{\alpha i} V^{}_{\beta j} V^{*}_{\alpha j} V^{*}_{\beta i} \; ,
\end{equation}
where the Greek and Latin subscripts run over $(e, \mu, \tau)$ and $(1, 2, 3)$, respectively.  $^{\alpha i}_{}\Box^{}_{\beta j}$ are not automatically real if $\alpha \neq \beta$ and $i \neq j$. In this paper, no summation on repeated indices is implied. The imaginary parts of $^{\alpha i}_{}\Box^{}_{\beta j}$
\begin{equation}
^{\alpha i}_{}\Im^{}_{\beta j} \; \equiv \; {\rm Im} ^{\alpha i}_{}\Box^{}_{\beta j} \; \equiv \; {\rm Im} \left [ V^{}_{\alpha i} V^{}_{\beta j} V^{*}_{\alpha j} V^{*}_{\beta i} \right ] \; ,
\end{equation}
are the measures of the CP violation in $V$. For three fermion generations, there is only one such independent complex quantity, all $^{\alpha i}_{}\Im^{}_{\beta j}$ equal to the Jarlskog invariant ${\cal J}$ except a sign difference. Correspondingly we define the real parts of $^{\alpha i}_{}\Box^{}_{\beta j}$ as
\begin{equation}
^{\alpha i}_{}\Re^{}_{\beta j} \; \equiv \; {\rm Re} ^{\alpha i}_{}\Box^{}_{\beta j} \; \equiv \; {\rm Re} \left [ V^{}_{\alpha i} V^{}_{\beta j} V^{*}_{\alpha j} V^{*}_{\beta i} \right ] \; .
\end{equation}
Neutrino oscillation probabilities are linear in $^{\alpha i}_{}\Re^{}_{\beta j}$ and $^{\alpha i}_{}\Im^{}_{\beta j}$, enabling a straightforward description of oscillation data.
\begin{eqnarray}
P(\nu^{}_{\alpha} \rightarrow \nu^{}_{\beta}) & = & \delta^{}_{\alpha \beta} - 4 \sum^{}_{j > i} {^{\alpha i}_{}\Re^{}_{\beta j}} \sin^2 \frac{\Delta m^{2}_{ji} L}{4 E} + 2 \sum^{}_{j > i} {^{\alpha i}_{}\Im^{}_{\beta j}} \sin \frac{\Delta m^{2}_{ji} L}{4 E} \nonumber\\
& = & \delta^{}_{\alpha \beta} - 4 \sum^{}_{j > i} {^{\alpha i}_{}\Re^{}_{\beta j}} \sin^2 \frac{\Delta m^{2}_{ji} L}{4 E} \nonumber\\
& & \; + \; 8 \; {\cal J} \sum^{}_{\gamma} \epsilon^{}_{\alpha\beta\gamma} \sin\frac{\Delta m^{2}_{21} L}{4 E} \sin\frac{\Delta m^{2}_{31} L}{4 E} \sin\frac{\Delta m^{2}_{32} L}{4 E} \; .
\end{eqnarray}
General analyses of neutrino oscillations among three flavors can readily determine the boxes \cite{oscillation}.

Enlightened by the idea of the $3\times 3$ CKM angle matrix proposed by Harrison {\it et al.} \cite{Harrison:2009bz}, we can similarly construct the $3\times 3$ MNS angle matrix $\Phi$ for Dirac neutrinos
\begin{equation}
\Phi \; = \; \left ( \begin{matrix} \Phi^{}_{e1} & ~\Phi^{}_{e2}~ & \Phi^{}_{e3} \cr \Phi^{}_{\mu 1} & \Phi^{}_{\mu 2} & \Phi^{}_{\mu 3} \cr \Phi^{}_{\tau 1} & \Phi^{}_{\tau 2} & \Phi^{}_{\tau 3} \end{matrix} \right ) \; ,
\end{equation}
by using the angles of nine box invariants

\begin{equation}
\Phi^{}_{\alpha i} \; \equiv \; - \arg \left(- ^{\beta j}_{}\Box^{}_{\gamma k} \right) \; ,
\end{equation}
where $\alpha$, $\beta$ and $\gamma$ run co-cyclically over $e$, $\mu$ and $\tau$, while $i$, $j$ and $k$ run co-cyclically over $1$, $2$ and $3$. 

We can easily find that the absolute value of $\Phi^{}_{\alpha i}$ is just the inner angle shared by the UTs $\triangle^{}_{\alpha}$ and $\triangle^{}_{i}$ as one can see in Fig. 1. Besides, the common sign of all nine angle matrix elements is just the sign of the Jarlskog invariant ${\cal J}$. If ${\cal J} > 0$, all the nine angles in  $\Phi$ lie between $0$ and $\pi$; if ${\cal J} < 0$, all the nine angles lie between $- \pi$ and $0$
\footnote{There is another immediate way to find out the sign of ${\cal J}$. We firstly define the sequence of the sides of the UTs:  for triangles $\triangle^{}_{e}$, $\triangle^{}_{\mu}$ and $\triangle^{}_{\tau}$, we follow the sequence of $V^{}_{\alpha 1} V^{*}_{\beta 1} \rightarrow V^{}_{\alpha 2} V^{*}_{\beta 2} \rightarrow V^{}_{\alpha 3} V^{*}_{\beta 3}$, where $\alpha \beta = \mu \tau$, $\tau e$ and $e \mu$ respectively; and for triangles $\triangle^{}_{1}$, $\triangle^{}_{2}$ and $\triangle^{}_{3}$, we follow the sequence of $V^{}_{e i} V^{*}_{e j} \rightarrow V^{}_{\mu i} V^{*}_{\mu j} \rightarrow V^{}_{\tau i} V^{*}_{\tau j}$, where $i j = 2 3$, $3 1$ and $1 2$ respectively. Then  all the UTs can be sorted into two classes: clockwise triangles and anti-clockwise triangles. We can easily find that if ${\cal J} > 0$, all six UTs (as shown in Fig. 1) are clockwise triangles while a negative ${\cal J}$ indicates that all six UTs are anti-clockwise triangles. This rule is true for both the Dirac and the Majorana neutrinos.}.
Each row or column of $\Phi$ corresponds to one UT, and the unitarity of $V$ now implies that elements of $\Phi$ satisfy the normalization conditions
\begin{equation}
\sum_\alpha \Phi^{}_{\alpha i} \; = \; \sum_i \Phi^{}_{\alpha i} \; = \; \pi \; .
\end{equation}
We can draw from Eq. (11) that there are only four independent real parameters in $\Phi$, same as the number of the independent real parameters in the unitary MNS matrix $V$. We can further prove that in the presence of CP violation, the angle matrix is fully equivalent to the MNS matrix. In Appendix A, we show how to re-obtain the mixing matrix $V$ from the angle matrix $\Phi$, the process is analogy to that in the quark sector \cite{Harrison:2009bz}.

The latest global analysis of current neutrino oscillation data yields $0.27 < \sin^2\theta^{}_{12} < 0.36$, $0.39 < \sin^2\theta^{}_{23} < 0.64$ and $0.001 < \sin^2\theta^{}_{13} < 0.035$ (NH) or $0.001 < \sin^2\theta^{}_{13} < 0.039$ (IH) at the $3 \sigma$ level \cite{Global Fit}, where ``NH" and ``IH" correspond to the normal and inverted neutrino mass hierarchies respectively. The CP-violating phases remain totally unconstrained. Correspondingly, the allowed range of the moduli of the elements of the MNS matrix $|V^{}_{\alpha i}|$ and the elements of the Dirac angle matrix $\Phi^{}_{\alpha i}$ can be obtained:
\begin{equation}
|V| \; = \; \left ( \begin{matrix} ~~ \begin{matrix} 0.786 \sim 0.854 \cr 0.784 \sim 0.854 \end{matrix} ~~ & ~~ \begin{matrix} 0.510 \sim 0.600 \cr 0.510 \sim 0.600 \end{matrix} ~~ & ~~ \begin{matrix} 0.032 \sim 0.187 \cr 0.032 \sim 0.197 \end{matrix} ~~ \\[6mm] ~~ \begin{matrix} 0.184 \sim 0.562 \cr 0.177 \sim 0.567 \end{matrix} ~~ & ~~ \begin{matrix} 0.390 \sim 0.728 \cr 0.385 \sim 0.731 \end{matrix} ~~ & ~~ \begin{matrix} 0.613 \sim 0.800 \cr 0.612 \sim 0.800 \end{matrix} ~~ \\[6mm] ~~ \begin{matrix} 0.200 \sim 0.570 \cr 0.193 \sim 0.575 \end{matrix} ~~ & ~~ \begin{matrix} 0.412 \sim 0.742 \cr 0.407 \sim 0.745 \end{matrix} ~~ & ~~ \begin{matrix} 0.589 \sim 0.781 \cr 0.588 \sim 0.781 \end{matrix} ~~ \end{matrix} \right ) \; ,
\end{equation}
\begin{equation}
\Phi \; = \; \left ( \begin{matrix} ~~ \begin{matrix} -16.7^{\circ} \sim 16.7^{\circ} \cr -17.6^{\circ} \sim 17.6^{\circ} \end{matrix} ~~ & ~~ \begin{matrix} -35.8^{\circ} \sim 35.8^{\circ} \cr -37.7^{\circ} \sim 37.7^{\circ} \end{matrix} ~~ & ~~~ \begin{matrix} 131.2^{\circ} \sim 228.8^{\circ} \cr 128.6^{\circ} \sim 231.4^{\circ} \end{matrix} ~~ \\[6mm] ~ -180^{\circ} \sim 180^{\circ} ~~ & ~ -180^{\circ} \sim 180^{\circ} ~~ & ~ \begin{matrix} -30.0^{\circ} \sim 30.0^{\circ} \cr -31.6^{\circ} \sim 31.6^{\circ} \end{matrix} ~~ \\[6mm] ~ -180^{\circ} \sim 180^{\circ} ~~ & ~ -180^{\circ} \sim 180^{\circ} ~~ & ~ \begin{matrix} -31.9^{\circ} \sim 31.9^{\circ} \cr -33.7^{\circ} \sim 33.7^{\circ} \end{matrix} ~~ \end{matrix} \right ) \; ,
\end{equation}
where the upper (lower) row corresponds to normal (inverted) neutrino mass hierarchy. The Jarlskog invariant ${\cal J}$ can range between $[-0.0433, 0.0433]$ (NH) or $[-0.0455, 0.0455]$ (IH) at the $3 \sigma$ level. 

The question is more pressing when we consider the case of Majorana neutrinos. Due to the Majorana nature of the neutrinos, the phases of three neutrino fields in Eq. (3) can not be freely chosen. The phase rotations on the charged lepton fields lead to the redefinition
\begin{equation}
V \rightarrow e^{i \Phi^{}_{l}} V \; ,
\end{equation}
where $\Phi^{}_{l} = {\rm diag} (\phi^{}_{l^{}_{1}}, \phi^{}_{l^{}_{2}}, \phi^{}_{l^{}_{3}})$. Therefore, for Majorana neutrinos, we have the new rephasing invariants \cite{Majorana CP}
\begin{equation}
S^{}_{\alpha i j} \; \equiv \; V^{}_{\alpha i} V^{*}_{\alpha j} \; ,
\end{equation}
which are not quartic but quadric products of $V^{}_{\alpha i}$ and are not obviously real if $i \neq j$.

In order to include the informations of Majorana-type CP violation, we introduce the following $3 \times 3$ Majorana angle matrix
\begin{equation}
\Psi \; = \; \left ( \begin{matrix} \Psi^{}_{e1} & ~\Psi^{}_{e2}~ & \Psi^{}_{e3} \cr \Psi^{}_{\mu 1} & \Psi^{}_{\mu 2} & \Psi^{}_{\mu 3} \cr \Psi^{}_{\tau 1} & \Psi^{}_{\tau 2} & \Psi^{}_{\tau 3} \end{matrix} \right ) \; ,
\end{equation}
with its nine elements are the angles of $S$ parameters in Eq. (15)
\begin{equation}
\Psi^{}_{\alpha i} \; \equiv \; \arg S^{}_{\alpha j k} \; = \; \arg \left( V^{}_{\alpha j} V^{*}_{\alpha k} \right) \; ,
\end{equation}
where $i$, $j$ and $k$ run co-cyclically over $1$, $2$ and $3$.  We can find that the three matrix elements in each row of $\Psi$ sum to zero:
\begin{equation}
\sum_i \Psi^{}_{\alpha i} \; = \; \arg \left ( \prod_i |V^{}_{\alpha i}|^2 \right ) \; = \; 0 \; .
\end{equation}
Above normalization conditions are satisfied independent of if $V$ is unitary. Then the number of independent real parameters in $\Psi$ is six, equals to the number of free parameters in the Majorana neutrino mixing matrix. In the case of Majorana neutrinos, we can also reconstruct the leptonic mixing matrix $V$ from the Majorana angle matrix $\Psi$, the detail processes can be found in Appendix A. 

We can easily see from Eqs. (5) and (15) that, there are the constitutive relations
\begin{equation}
^{\alpha i}_{}\Box^{}_{\beta j} \; = \; S^{}_{\alpha i j} S^{*}_{\beta i j} \; .
\end{equation}
Therefore we can easily construct the $\Phi$-matrix from the $\Psi$-matrix:
\begin{equation}
\Phi^{}_{\alpha i} \; = \; \Psi^{}_{\gamma i} - \Psi^{}_{\beta i} + \pi \; ,
\end{equation}
where $\alpha$, $\beta$ and $\gamma$ run co-cyclically over $e$, $\mu$ and $\tau$. It means we can also write out the Dirac angle matrix $\Phi$ for Majorana neutrinos, but are unable to draw informations about the Majorana phases from it.

From Eq. (19) we can find that even all $^{\alpha i}_{}\Im^{}_{\beta j}$ are zero (i.e., ${\cal J} = 0$ and no Dirac-type CP violation in $V$), the imaginary part of some $S^{}_{\alpha i j}$ can be nonzero and stands for the CP violation in $V$. On the contrary, if $S^{}_{\alpha i j}$ are all real, there is no CP violation in $V$ and ${\cal J} $ also equals to zero. It means that Dirac-type CP violation requires the existence of the Majorana-type CP violation but obviously the converse is not true \cite{Majorana CP}.

Here we introduce another combination of $S^{}_{\alpha i j}$
\begin{equation}
\eta^{}_{\alpha \beta i j} \; \equiv \; S^{}_{\alpha ij} S^{}_{\beta i j} \; = \; V^{}_{\alpha i} V^{*}_{\alpha j} V^{}_{\beta i} V^{*}_{\beta j} \; ,
\end{equation}
in which informations of the Majorana-type CP violation are also involved. For the sake of concision, we define the following notations
\begin{equation}
{\mathbb R}^{}_{\alpha \beta i j} \; \equiv \; {\rm Re} \; \eta^{}_{\alpha \beta i j} ~~~ {\rm and} ~~~ {\mathbb I}^{}_{\alpha \beta i j} \; \equiv \; {\rm Im} \; \eta^{}_{\alpha \beta i j}  \; .
\end{equation}
The parameters $\Psi^{}_{\alpha i}$, $S^{}_{\alpha i j}$ and $\eta^{}_{\alpha \beta i j}$ can show up in a variety of lepton number violating processes including $0 \nu \beta \beta$ decay and possibly leptogenesis {\it et al}. For example, the effective neutrino mass $\langle m \rangle^{}_{ee}$ in $0 \nu \beta \beta$ decay can be expressed as
\begin{equation}
\langle m \rangle^{}_{ee} \; = \; \left | \sum^{}_{i} V^{2}_{e i} \; m^{}_{i} \right | \; = \; \left | |V^{}_{e1}|^2 e^{- 2 i \Psi^{}_{e 2}} m^{}_{1} + |V^{}_{e2}|^2 e^{- 2 i \Psi^{}_{e 1}} m^{}_{2} +|V^{}_{e3}|^2 m^{}_{3} \right | \; .
\end{equation}

Another point worth to mention is that the nine elements of $\Psi$ are physically related to the orientations of the nine sides of UTs $\triangle^{}_{1}$, $\triangle^{}_{2}$ and $\triangle^{}_{3}$ in the complex plane (see Fig. 1). Therefore, the orientations of these three UTs have physical meanings if neutrinos are Majorana particles \cite{Majorana UTs}. On the other side, if the neutrinos are Dirac particles, the orientations of the UTs have no physical meaning, reflecting the fact that Dirac UTs rotate under rephasing of the charged lepton fields or the neutrino fields.

Before our analysis of the radiative corrections to the angle matrix $\Phi$ and $\Psi$, we show here that the conceptions of the Dirac and the Majorana angle matrices can be extrapolated to account for any generation of neutrinos. There is no a unique way to choose the elements of $\Phi^{(N)}_{}$ and $\Psi^{(N)}_{}$ for $N$ ($N > 3$) generation neutrinos, we give one possible choice here. Here ($e$, $\mu$, $\tau$, $s_4$, $s_5$, $\dots$, $s_N$) stand for the flavor indices and ($1$, $2$, $3$, $4$, $5$, $\dots$, $N$) for the mass indices. 

For $N$ generation of Dirac neutrinos,  we define the $N \times N$ Dirac angle matrix
\begin{eqnarray}
\Phi^{(N)}_{} & = & \left ( \begin{matrix} \Phi^{}_{e1} & \Phi^{}_{e2} & \Phi^{}_{e3} & \cdots & \Phi^{}_{eN} \cr \Phi^{}_{\mu 1} & \Phi^{}_{\mu 2} & \Phi^{}_{\mu 3} & \cdots & \Phi^{}_{\mu N} \cr \Phi^{}_{\tau 1} & \Phi^{}_{\tau 2} & \Phi^{}_{\tau 3} & \cdots & \Phi^{}_{\tau N} \cr \vdots & \vdots & \vdots & \ddots & \vdots \cr \Phi^{}_{s_N 1} & \Phi^{}_{s_N 2} & \Phi^{}_{s_N 3} & \cdots & \Phi^{}_{s_N N} \end{matrix} \right ) \nonumber\\ \nonumber\\
& = & \left ( \begin{matrix} - \arg \left(- ^{\mu 2}_{}\Box^{}_{\tau 3} \right) & - \arg \left(- ^{\mu 3}_{}\Box^{}_{\tau 4} \right) & - \arg \left(- ^{\mu 4}_{}\Box^{}_{\tau 5} \right) & \cdots & - \arg \left(- ^{\mu 1}_{}\Box^{}_{\tau 2} \right) \cr - \arg \left(- ^{\tau 2}_{}\Box^{}_{s_4 3} \right) & - \arg \left(- ^{\tau 3}_{}\Box^{}_{s_4 4} \right) & - \arg \left(- ^{\tau 4}_{}\Box^{}_{s_4 5} \right) & \cdots & - \arg \left(- ^{\tau 1}_{}\Box^{}_{s_4 2} \right) \cr - \arg \left(- ^{s_4 2}_{}\Box^{}_{s_5 3} \right) & - \arg \left(- ^{s_4 3}_{}\Box^{}_{s_5 4} \right) & - \arg \left(- ^{s_4 4}_{}\Box^{}_{s_5 5} \right) & \cdots & - \arg \left(- ^{s_4 1}_{}\Box^{}_{s_5 2} \right) \cr \vdots & \vdots & \vdots & \ddots & \vdots \cr - \arg \left(- ^{e 2}_{}\Box^{}_{\mu 3} \right) & - \arg \left(- ^{e 3}_{}\Box^{}_{\mu 4} \right) & - \arg \left(- ^{e 4}_{}\Box^{}_{\mu 5} \right) & \cdots & - \arg \left(- ^{e 1}_{}\Box^{}_{\mu 2} \right) \end{matrix} \right ) \; , \nonumber\\
\end{eqnarray}
and its elements $\Phi^{(N)}_{\alpha i}$ satisfy the normalization conditions
\begin{equation}
\sum_\alpha \Phi^{(N)}_{\alpha i} \; = \; \sum_i \Phi^{(N)}_{\alpha i} \; = \; \left ( N - 2 \right ) \pi \; ,
\end{equation}
where $\alpha$ stands for the flavor index and $i$ for the mass index. Then there are altogether $\displaystyle \frac{1}{2} \left ( N - 1 \right ) \left ( N - 2 \right )$ independent parameters in $\Phi^{(N)}_{}$ which is equivalent to the number of the independent real parameters in a  $N \times N$ unitary mixing matrix.

For $N$ generation of Majorana neutrinos, the $N \times N$ Majorana angle matrix can be defined as
\begin{eqnarray}
\Psi^{(N)}_{} & = & \left ( \begin{matrix} \Psi^{}_{e1} & \Psi^{}_{e2} & \Psi^{}_{e3} & \cdots & \Psi^{}_{eN} \cr \Psi^{}_{\mu 1} & \Psi^{}_{\mu 2} & \Psi^{}_{\mu 3} & \cdots & \Psi^{}_{\mu N} \cr \Psi^{}_{\tau 1} & \Psi^{}_{\tau 2} & \Psi^{}_{\tau 3} & \cdots & \Psi^{}_{\tau N} \cr \vdots & \vdots & \vdots & \ddots & \vdots \cr \Psi^{}_{s_N 1} & \Psi^{}_{s_N 2} & \Psi^{}_{s_N 3} & \cdots & \Psi^{}_{s_N N} \end{matrix} \right ) \nonumber\\ \nonumber\\
& = & \left ( \begin{matrix} \arg \left( V^{}_{e 2} V^{*}_{e 3} \right) & \arg \left( V^{}_{e 3} V^{*}_{e 4} \right) & \arg \left( V^{}_{e 4} V^{*}_{e 5} \right) & \cdots & \arg \left( V^{}_{e 1} V^{*}_{e 2} \right) \cr \arg \left( V^{}_{\mu 2} V^{*}_{\mu 3} \right) & \arg \left( V^{}_{\mu 3} V^{*}_{\mu 4} \right) & \arg \left( V^{}_{\mu 4} V^{*}_{\mu 5} \right) & \cdots & \arg \left( V^{}_{\mu 1} V^{*}_{\mu 2} \right) \cr \arg \left( V^{}_{\tau 2} V^{*}_{\tau 3} \right) & \arg \left( V^{}_{\tau 3} V^{*}_{\tau 4} \right) & \arg \left( V^{}_{\tau 4} V^{*}_{\tau 5} \right) & \cdots & \arg \left( V^{}_{\tau 1} V^{*}_{\tau 2} \right) \cr \vdots & \vdots & \vdots & \ddots & \vdots \cr \; \arg \left( V^{}_{s_N 2} V^{*}_{s_N 3} \right) \; & \; \arg \left( V^{}_{s_N 3} V^{*}_{s_N 4} \right) \; & \; \arg \left( V^{}_{s_N 4} V^{*}_{s_N 5} \right) \; & \; \cdots \; & \; \arg \left( V^{}_{s_N 1} V^{*}_{s_N 2} \right) \; \end{matrix} \right ) \; , \nonumber\\
\end{eqnarray}
and the $N$ matrix elements in each row of $\Psi^{(N)}_{}$ satisfy the normalization conditions
\begin{equation}
\sum_i \Psi^{}_{\alpha i} \; = \; \arg \left ( \prod_i |V^{}_{\alpha i}|^2 \right ) \; = \; 0 \; ,
\end{equation}
where $\alpha$ stands for the flavor index and $i$ for the mass index. There are totally $\displaystyle \frac{1}{2} N \left ( N - 1 \right )$ independent real parameters in $\Psi^{(N)}_{}$. Again, this number is equivalent to that of the $N \times N$ Majorana neutrino mixing matrix. In case of $N$ generation Majorana neutrinos, we can also construct the $\Phi$ matrix from the $\Psi$ matrix:
\begin{equation}
\Phi^{}_{\alpha i} \; = \; \Psi^{}_{\gamma i} - \Psi^{}_{\beta i} + \pi \; ,
\end{equation}
where $\beta$ and $\gamma$ are the next two flavor indices right after $\alpha$.

\section{One-Loop Renormalization Group Equations}

The RGEs of neutrino masses and mixing have been discussed in many papers with variety of parametrizations \cite{Majorana v.s. Dirac, RGE, phases RGE, Dirac RGE, Majorana RGE}. It has been shown that three neutrino masses and the mixing matrix may receive large radiative corrections, especially if neutrino masses are nearly degenerate or in case of the MSSM with large $\tan\beta$. Studies also show that the running behaviours can be quite different for Dirac or Majorana neutrinos \cite{Majorana v.s. Dirac}, and the additional Majorana phases may have intrinsic behaviour in the evolution \cite{phases RGE}. In this section, we proceed to consider the one-loop RGEs of $|V^{}_{\alpha i}|^2$, $\Phi$, $\Psi$ and ${\cal J}$ for both the Dirac and the Majorana neutrinos. Note that all these parameters are rephasing invariant and independent of any particular parametrization of the MNS matrix $V$. Distinguishable RGE running effects between Dirac neutrinos and Majorana neutrinos are discussed in detail.

\subsection{Dirac Neutrinos}

If neutrinos are Dirac particles, their Yukawa coupling matrix $Y^{}_{\nu}$ must be extremely suppressed in magnitude to reproduce the light neutrino masses of ${\cal O}(1)$ eV or smaller at low energy scales. The running of $Y^{}_{\nu}$ from the electroweak energy scale $\Lambda^{}_{\rm EW}$ to a superhigh energy scale $\Lambda$ is governed by the one-loop RGE \cite{Dirac RGE}
\begin{equation}
16\pi^2 \frac{{\rm d}\omega}{{\rm d}t} \; = \; 2 \alpha^{}_{\rm D} \omega + C^{}_{} \left [ \left ( Y^{}_{l} Y^{\dagger}_{l} \right ) \omega + \omega \left ( Y^{}_{l} Y^{\dagger}_{l} \right ) \right ] \; ,
\end{equation}
where $\omega \equiv Y^{}_{\nu} Y^{\dagger}_{\nu}$ is a Hermitian quantity, $t \equiv \ln \left ( \mu / \Lambda \right )$ with $\mu$ being an arbitrary renormalization scale between $\Lambda^{}_{\rm EW}$ and $\Lambda$, $Y^{}_{l}$ is the charged-lepton Yukawa coupling matrix, $C = -1.5$ (SM) or $C = 1$ (MSSM) and $\alpha^{}_{\rm D} \approx - 0.45 g^{2}_{1} - 2.25 g^{2}_{2} + 3 y^{2}_{t}$ (SM) or $\alpha^{}_{\rm D} \approx - 0.6 g^{2}_{1} - 3 g^{2}_{2} + 3 y^{2}_{t}$ (MSSM). Here $g^{}_{1}$ and $g^{}_{2}$ are the gauge couplings, $y^{}_{t}$ stands for the top-quark Yukawa coupling. In writing out Eq. (29), we have safely neglected those tiny terms of ${\cal O}(\omega^{2}_{})$.

Without loss of generality, we choose the flavor basis where $Y^{}_{l}$ is diagonal and real (positive): $Y^{}_{l} = {\rm diag} \left \{ y^{}_{e}, \; y^{}_{\mu}, \; y^{}_{\tau} \right \}$. In this basis $\omega$ can be diagonalized by the unitary transformation $V^{\dagger}_{} \omega V^{}_{} = \hat{\omega} \equiv {\rm diag} \left \{ y^{2}_{1}, \; y^{2}_{2}, \; y^{2}_{3} \right \}$, where $V$ is just the MNS matrix and at $\Lambda^{}_{\rm EW}$ the Dirac neutrino masses are $m^{}_{i} = v y^{}_{i}$ (SM) or $m^{}_{i} = v y^{}_{i} \sin\beta$ (MSSM) with $v \approx 174$ GeV. One may use Eq. (29) to derive the explicit RGEs of neutrino masses and the MNS matrix $V$. By taking some lengthy but not complicated calculations we can further derive the RGEs of those rephasing invariant parameters with the help of Eqs.(2), (5) and (10). Here we simply give the resulting one-loop RGEs of $|V^{}_{\alpha i}|^2$, $\Phi$ and ${\cal J}$. During the derivation, we have taken the approximation of $\tau$-lepton dominance. In other words, the contributions of $y^{2}_{e}$ and $y^{2}_{\mu}$ to all of the RGEs are negligibly small and are safely negligted.

\begin{eqnarray}
& & 16\pi^2 \frac{\rm d}{{\rm d}t} \left ( \begin{matrix} |V^{}_{e1}|^2 & |V^{}_{e2}|^2 & |V^{}_{e3}|^2 \cr |V^{}_{\mu 1}|^2 & |V^{}_{\mu 2}|^2 & |V^{}_{\mu 3}|^2 \cr |V^{}_{\tau 1}|^2 & |V^{}_{\tau 2}|^2 & |V^{}_{\tau 3}|^2 \cr \end{matrix} \right ) \nonumber \\ \nonumber \\
& & \; = \; 2 C^{}_{} y^2_\tau \left \{ \frac{m^{2}_{2} + m^{2}_{1}}{\Delta m^{2}_{21}} \left ( \begin{matrix} \; - {^{e1}_{}\Re^{}_{\tau 2}} & ~ ^{e1}_{}\Re^{}_{\tau 2} ~  & ~ 0 ~ \cr \; - {^{\mu1}_{}\Re^{}_{\tau 2}} & ~ ^{\mu1}_{}\Re^{}_{\tau 2} ~ & ~ 0 ~ \cr \; - {^{\tau1}_{}\Re^{}_{\tau 2}} & ~ ^{\tau1}_{}\Re^{}_{\tau 2} ~ & ~ 0 ~ \cr \end{matrix} \right ) + \; \frac{m^{2}_{3} + m^{2}_{1}}{\Delta m^{2}_{31}} \left ( \begin{matrix} \; - {^{e1}_{}\Re^{}_{\tau 3}} & ~ 0 ~ & ^{e1}_{}\Re^{}_{\tau 3} ~  \cr \; - {^{\mu1}_{}\Re^{}_{\tau 3}} & ~ 0 ~ & ^{\mu1}_{}\Re^{}_{\tau 3} ~ \cr \; - {^{\tau1}_{}\Re^{}_{\tau 3}} & ~ 0 ~ & ^{\tau1}_{}\Re^{}_{\tau 3} ~ \cr \end{matrix} \right ) \right. \nonumber \\ \nonumber \\
& & ~~~~~~~~~~~~~~ \left. + \; \frac{m^{2}_{3} + m^{2}_{2}}{\Delta m^{2}_{32}} \left ( \begin{matrix} ~ 0 & \; - {^{e2}_{}\Re^{}_{\tau 3}} ~ & ^{e2}_{}\Re^{}_{\tau 3} ~  \cr ~ 0 & \; - {^{\mu2}_{}\Re^{}_{\tau 3}} ~ & ^{\mu2}_{}\Re^{}_{\tau 3} ~ \cr ~ 0 & \; - {^{\tau2}_{}\Re^{}_{\tau 3}} ~ & ^{\tau2}_{}\Re^{}_{\tau 3} ~ \cr \end{matrix} \right ) \right \} \; , 
\end{eqnarray}

\begin{eqnarray}
& & 16\pi^2 \frac{\rm d}{{\rm d}t} \left ( \begin{matrix} \Phi^{}_{e1} & \Phi^{}_{e2} & \Phi^{}_{e3} \cr \Phi^{}_{\mu 1} & \Phi^{}_{\mu 2} & \Phi^{}_{\mu 3} \cr \Phi^{}_{\tau 1} & \Phi^{}_{\tau 2} & \Phi^{}_{\tau 3} \cr \end{matrix} \right ) \nonumber \\ \nonumber \\
& & = \; C^{}_{} y^2_\tau \; {\cal J} \left \{ \frac{m^{2}_{2} + m^{2}_{1}}{\Delta m^{2}_{21}} \left ( \begin{matrix} - |V^{}_{\mu 2}|^{-2} & \; |V^{}_{\mu 1}|^{-2} & |V^{}_{\mu 2}|^{-2} - |V^{}_{\mu 1}|^{-2} \\[2.1mm] - |V^{}_{e 2}|^{-2} & \; |V^{}_{e 1}|^{-2} & |V^{}_{e 2}|^{-2} - |V^{}_{e 1}|^{-2} \\[2.1mm] \; \displaystyle \frac{1 - |V^{}_{\tau 2}|^2}{|V^{}_{e 2}|^2 |V^{}_{\mu 2}|^2} ~ & ~ \displaystyle \frac{|V^{}_{\tau 1}|^2 - 1}{|V^{}_{e 1}|^2 |V^{}_{\mu 1}|^2} ~ & ~ \displaystyle \frac{1 - |V^{}_{\tau 1}|^2}{|V^{}_{e 1}|^2 |V^{}_{\mu 1}|^2} + \frac{|V^{}_{\tau 2}|^2 - 1}{|V^{}_{e 2}|^2 |V^{}_{\mu 2}|^2} \; \cr \end{matrix} \right ) \right. \nonumber \\ \nonumber \\
& & ~~~~~~~~~~~~~~~ \left. + \; \frac{m^{2}_{3} + m^{2}_{1}}{\Delta m^{2}_{31}} \left ( \begin{matrix} - |V^{}_{\mu 3}|^{-2} & |V^{}_{\mu 3}|^{-2} - |V^{}_{\mu 1}|^{-2} & ~ |V^{}_{\mu 1}|^{-2} \\[2.1mm] - |V^{}_{e 3}|^{-2} & |V^{}_{e 3}|^{-2} - |V^{}_{e 1}|^{-2} & ~ |V^{}_{e 1}|^{-2} \\[2.1mm] \; \displaystyle \frac{1 - |V^{}_{\tau 3}|^2}{|V^{}_{e 3}|^2 |V^{}_{\mu 3}|^2} ~ & ~ \displaystyle \frac{1 - |V^{}_{\tau 1}|^2}{|V^{}_{e 1}|^2 |V^{}_{\mu 1}|^2} + \frac{|V^{}_{\tau 3}|^2 - 1}{|V^{}_{e 3}|^2 |V^{}_{\mu 3}|^2} ~ & ~ \displaystyle \frac{|V^{}_{\tau 1}|^2 - 1}{|V^{}_{e 1}|^2 |V^{}_{\mu 1}|^2} \; \cr \end{matrix} \right ) \right. \nonumber \\ \nonumber \\
& & ~~~~~~~~~~~~~~~ \left. + \; \frac{m^{2}_{3} + m^{2}_{2}}{\Delta m^{2}_{32}} \left ( \begin{matrix} |V^{}_{\mu 3}|^{-2} - |V^{}_{\mu 2}|^{-2} & - |V^{}_{\mu 3}|^{-2} & ~ |V^{}_{\mu 2}|^{-2} \\[2.1mm] |V^{}_{e 3}|^{-2} - |V^{}_{e 2}|^{-2} & - |V^{}_{e 3}|^{-2} & ~ |V^{}_{e 2}|^{-2} \\[2.1mm] \; \displaystyle \frac{1 - |V^{}_{\tau 2}|^2}{|V^{}_{e 2}|^2 |V^{}_{\mu 2}|^2} + \frac{|V^{}_{\tau 3}|^2 - 1}{|V^{}_{e 3}|^2 |V^{}_{\mu 3}|^2} ~ & ~ \displaystyle \frac{1 - |V^{}_{\tau 3}|^2}{|V^{}_{e 3}|^2 |V^{}_{\mu 3}|^2} ~ & ~ \displaystyle \frac{|V^{}_{\tau 2}|^2 - 1}{|V^{}_{e 2}|^2 |V^{}_{\mu 2}|^2} \; \cr \end{matrix} \right ) \right \} \; , \nonumber \\
\end{eqnarray}
\begin{eqnarray}
16\pi^2 \frac{\rm d}{{\rm d}t} \; {\cal J} & = & C^{}_{} y^2_\tau \; {\cal J} \left \{ \frac{m^{2}_{2} + m^{2}_{1}}{\Delta m^{2}_{21}} \left ( |V^{}_{\tau 1}|^2 - |V^{}_{\tau 2}|^2 \right ) + \; \frac{m^{2}_{3} + m^{2}_{1}}{\Delta m^{2}_{31}} \left ( |V^{}_{\tau 1}|^2 - |V^{}_{\tau 3}|^2 \right ) \right. \nonumber \\[2.4mm]
& & ~~~ \left. + \; \frac{m^{2}_{3} + m^{2}_{2}}{\Delta m^{2}_{32}} \left ( |V^{}_{\tau 2}|^2 - |V^{}_{\tau 3}|^2 \right ) \right \} \; .
\end{eqnarray}

Since we have $\Delta m^{2}_{21} \ll \Delta m^{2}_{31} \approx \Delta m^{2}_{32}$, we can infer from Eq. (30) that in the standard parametrization $\theta^{}_{12} = \arctan (|V^{}_{e 2}| / |V^{}_{e 1}|)$ is in general more sensitive to the radiative correction than the other two mixing angles $\theta^{}_{13} = \arcsin (|V^{}_{e 3}|)$ and $\theta^{}_{23} = \arctan (|V^{}_{\mu 3}| / |V^{}_{\tau 3}|)$. This result is also true if the neutrinos are Majorana particles which can be easily seen from Eq. (41) and is consist with the analyses in many previous papers \cite{RGE, Dirac RGE, Majorana RGE}. 

Since the absolute mass scale of three neutrinos and the sign of $\Delta m^{2}_{31}$ remain unknown, we further consider three typical patterns of the neutrino mass spectrum: normal hierarchy (NH), inverted hierarchy (IH) and near degeneracy (ND).
\\

\begin{itemize}

\item {\bf Normal Hierarchy} ~ $m^{}_{3} \gg m^{}_{2} \gg m^{}_{1} \simeq 0$, $m^{}_{2} \approx \sqrt{\Delta m^{2}_{21}}$ and $m^{}_{3} \approx \sqrt{\Delta m^{2}_{31}}$

In this neutrino masses limit the one-loop RGEs of $\Phi$ and ${\cal J}$ can be approximately expressed as
\begin{eqnarray}
16\pi^2 \frac{\rm d}{{\rm d}t} \left ( \begin{matrix} \Phi^{}_{e1} & \Phi^{}_{e2} & \Phi^{}_{e3} \cr \Phi^{}_{\mu 1} & \Phi^{}_{\mu 2} & \Phi^{}_{\mu 3} \cr \Phi^{}_{\tau 1} & \Phi^{}_{\tau 2} & \Phi^{}_{\tau 3} \cr \end{matrix} \right ) & \approx & 2 C^{}_{} y^2_\tau \; {\cal J} \left \{ \frac{1}{|V^{}_{e 2}|^2 |V^{}_{\mu 2}|^2} \left ( \begin{matrix} - |V^{}_{e 2}|^2 & 0 & |V^{}_{e 2}|^2 \cr - |V^{}_{\mu 2}|^2 & 0 & |V^{}_{\mu 2}|^2 \cr \; 1 - |V^{}_{\tau 2}|^2 \; & ~ 0 ~ & \; |V^{}_{\tau 1}|^2 - 1 \; \cr \end{matrix} \right ) \right.\nonumber\\ \nonumber\\
& & ~~~~~~~~~~~~~~~ \left. + \; \frac{\Delta m^{2}_{21}}{\Delta m^{2}_{31}} \; \frac{1}{|V^{}_{e 3}|^2} \left ( \begin{matrix} \; 0 & \; 0 & \; 0 ~ \cr \; 1& -1 & \; 0 ~ \cr -1 & \; 1 & \; 0 ~ \cr \end{matrix} \right ) \; \right \} \; , 
\end{eqnarray}
\begin{equation}
16\pi^2 \frac{\rm d}{{\rm d}t} \; {\cal J} \; \approx \; 2 C^{}_{} y^2_\tau \; {\cal J} \left ( |V^{}_{\tau 1}|^2 - |V^{}_{\tau 3}|^2 \right ) \; ,
\end{equation}
where in the next leading order terms lead by $\Delta m^{2}_{21} / \Delta m^{2}_{31}$, we preserve only those terms inversely proportional to $|V^{}_{e3}|$. Taking into account of the smallness of $|V^{}_{e3}|$, the contribution of these terms may be comparable with the leading order terms.

Some discussions are in order.
\begin{itemize}

\item The angle $\Phi^{}_{e2}$ is most insensitive to the radiative correction in the leading order if neutrino masses are of normal hierarchy.

\item In case of the MSSM with large $\tan \beta$, $\Delta m^{2}_{21} / \Delta m^{2}_{31}$ increases significantly with the increase of energy scale \cite{Dirac RGE}, therefore $\Phi^{}_{\mu 1}$, $\Phi^{}_{\mu 2}$, $\Phi^{}_{\tau 1}$ and $\Phi^{}_{\tau 2}$ are probably having significant evolutions, especially when $|V^{}_{e3}|$ takes a small value. While in the SM, all nine angles receive only small radiative corrections. 

\item The one-loop RGE of ${\cal J}$ is proportional to ${\cal J}$ itself, therefore its evolutions are apparently opposite for positive and negative ${\cal J}$. 

\item Consider the evolution of ${\cal J}$ from  $\Lambda^{}_{\rm EW}$ to a superhigh energy scale. We can find from Eq. (12) that $|V^{}_{\tau 1}|^2 - |V^{}_{\tau 3}|^2$ is negative at $\Lambda^{}_{\rm EW}$. In the SM, $C < 0$, we can expect that $|{\cal J}|$ only slightly increases during the evolution. In the MSSM, $C > 0$, and we can find from Eq. (30) that $|V^{}_{\tau}|^2$ increases with the evolution while $|V^{}_{\tau}|^2$ decreases. Therefore $|V^{}_{\tau 1}|^2 - |V^{}_{\tau 3}|^2$ keeps negative which indicates that $|{\cal J}|$ will go approaching zero during the evolution. 

\end{itemize}

\item {\bf Inverted Hierarchy}  ~ $\sqrt{ - \Delta m^{2}_{31}} \approx m^{}_{2} \approx m^{}_{1} \gg m^{}_{3} \approx 0$

In this neutrino masses limit the one-loop RGEs of $\Phi$ and ${\cal J}$ can be approximately expressed as
\begin{eqnarray}
& & 16\pi^2 \frac{\rm d}{{\rm d}t} \left ( \begin{matrix} \Phi^{}_{e1} & \Phi^{}_{e2} & \Phi^{}_{e3} \cr \Phi^{}_{\mu 1} & \Phi^{}_{\mu 2} & \Phi^{}_{\mu 3} \cr \Phi^{}_{\tau 1} & \Phi^{}_{\tau 2} & \Phi^{}_{\tau 3} \cr \end{matrix} \right ) \nonumber \\ \nonumber \\
& & \approx \; - C^{}_{} y^2_\tau \; {\cal J} \left \{ \frac{2 \Delta m^{2}_{31}}{\Delta m^{2}_{21}} \left ( \begin{matrix} - |V^{}_{\mu 2}|^{-2} & \; |V^{}_{\mu 1}|^{-2} & |V^{}_{\mu 2}|^{-2} - |V^{}_{\mu 1}|^{-2} \\[2.1mm] - |V^{}_{e 2}|^{-2} & \; |V^{}_{e 1}|^{-2} & |V^{}_{e 2}|^{-2} - |V^{}_{e 1}|^{-2} \\[2.1mm] \; \displaystyle \frac{1 - |V^{}_{\tau 2}|^2}{|V^{}_{e 2}|^2 |V^{}_{\mu 2}|^2} ~ & ~ \displaystyle \frac{|V^{}_{\tau 1}|^2 - 1}{|V^{}_{e 1}|^2 |V^{}_{\mu 1}|^2} ~ & ~ \displaystyle \frac{1 - |V^{}_{\tau 1}|^2}{|V^{}_{e 1}|^2 |V^{}_{\mu 1}|^2} + \frac{|V^{}_{\tau 2}|^2 - 1}{|V^{}_{e 2}|^2 |V^{}_{\mu 2}|^2} \; \cr \end{matrix} \right ) \right. \nonumber \\ \nonumber \\
& & ~~~~~~~ \left. + \; \left ( \begin{matrix} - |V^{}_{\mu 2}|^{-2} & - |V^{}_{\mu 1}|^{-2} & \displaystyle \frac{1 - |V^{}_{\mu 3}|^2}{|V^{}_{\mu 1}|^2 |V^{}_{\mu 2}|^2} \\[3.6mm] - |V^{}_{e 2}|^{-2} & - |V^{}_{e 1}|^{-2} & \displaystyle \frac{1 - |V^{}_{e 3}|^2}{|V^{}_{e 1}|^2 |V^{}_{e 2}|^2} \\[3.6mm] \; \displaystyle \frac{1 - |V^{}_{\tau 2}|^2}{|V^{}_{e 2}|^2 |V^{}_{\mu 2}|^2} ~ & ~ \displaystyle \frac{1 - |V^{}_{\tau 1}|^2}{|V^{}_{e 1}|^2 |V^{}_{\mu 1}|^2} ~ & - |V^{}_{e 1}|^{-2} - |V^{}_{e 2}|^{-2} - |V^{}_{\mu 1}|^{-2} - |V^{}_{\mu 2}|^{-2} \; \cr \end{matrix} \right ) \right \} \; , \nonumber \\
\end{eqnarray}
\begin{equation}
16\pi^2 \frac{\rm d}{{\rm d}t} \; {\cal J} \; \approx \; - C^{}_{} y^2_\tau \; {\cal J} \left [ \frac{2 \Delta m^{2}_{31}}{\Delta m^{2}_{21}} \left ( |V^{}_{\tau 1}|^2 - |V^{}_{\tau 2}|^2 \right ) + \left ( 1 - 3 |V^{}_{\tau 3}|^2 \right ) \right ]  \; .
\end{equation}

Note that, in the case of IH, $\Delta m^{2}_{31}$ is negative.

\item {\bf Near Degeneracy}  ~ $m^{}_{3} \approx m^{}_{2} \approx m^{}_{1} $ and $\Delta m^{2}_{32} \approx \Delta m^{2}_{31}$.

If three neutrino masses are nearly degenerate, the one-loop RGEs of $\Phi$ and ${\cal J}$ can be approximately expressed as

\begin{eqnarray}
& & 16\pi^2 \frac{\rm d}{{\rm d}t} \left ( \begin{matrix} \Phi^{}_{e1} & \Phi^{}_{e2} & \Phi^{}_{e3} \cr \Phi^{}_{\mu 1} & \Phi^{}_{\mu 2} & \Phi^{}_{\mu 3} \cr \Phi^{}_{\tau 1} & \Phi^{}_{\tau 2} & \Phi^{}_{\tau 3} \cr \end{matrix} \right ) \nonumber \\ \nonumber \\
& & \approx \; 2 C^{}_{} y^2_\tau \; {\cal J} \left \{ \frac{m^{2}_{1}}{\Delta m^{2}_{21}} \left ( \begin{matrix} - |V^{}_{\mu 2}|^{-2} & \; |V^{}_{\mu 1}|^{-2} & |V^{}_{\mu 2}|^{-2} - |V^{}_{\mu 1}|^{-2} \\[2.1mm] - |V^{}_{e 2}|^{-2} & \; |V^{}_{e 1}|^{-2} & |V^{}_{e 2}|^{-2} - |V^{}_{e 1}|^{-2} \\[2.1mm] \; \displaystyle \frac{1 - |V^{}_{\tau 2}|^2}{|V^{}_{e 2}|^2 |V^{}_{\mu 2}|^2} ~ & ~ \displaystyle \frac{|V^{}_{\tau 1}|^2 - 1}{|V^{}_{e 1}|^2 |V^{}_{\mu 1}|^2} ~ & ~ \displaystyle \frac{1 - |V^{}_{\tau 1}|^2}{|V^{}_{e 1}|^2 |V^{}_{\mu 1}|^2} + \frac{|V^{}_{\tau 2}|^2 - 1}{|V^{}_{e 2}|^2 |V^{}_{\mu 2}|^2} \; \cr \end{matrix} \right ) \right. \nonumber \\ \nonumber \\
& & \left. + \; \frac{m^{2}_{1}}{\Delta m^{2}_{31}} \left ( \begin{matrix} - |V^{}_{\mu 2}|^{-2} & - |V^{}_{\mu 1}|^{-2} & \displaystyle \frac{1 - |V^{}_{\mu 3}|^2}{|V^{}_{\mu 1}|^2 |V^{}_{\mu 2}|^2} \\[3.6mm] - |V^{}_{e 2}|^{-2} & - |V^{}_{e 1}|^{-2} & \displaystyle \frac{1 - |V^{}_{e 3}|^2}{|V^{}_{e 1}|^2 |V^{}_{e 2}|^2} \\[3.6mm] \; \displaystyle \frac{1 - |V^{}_{\tau 2}|^2}{|V^{}_{e 2}|^2 |V^{}_{\mu 2}|^2} ~ & ~ \displaystyle \frac{1 - |V^{}_{\tau 1}|^2}{|V^{}_{e 1}|^2 |V^{}_{\mu 1}|^2} ~ & - |V^{}_{e 1}|^{-2} - |V^{}_{e 2}|^{-2} - |V^{}_{\mu 1}|^{-2} - |V^{}_{\mu 2}|^{-2} \; \cr \end{matrix} \right ) \right \} \; , \nonumber \\
\end{eqnarray}
\begin{equation}
16\pi^2 \frac{\rm d}{{\rm d}t} \; {\cal J} \; \approx \; 2 C^{}_{} y^2_\tau \; {\cal J} \left [ \frac{m^{2}_{1}}{\Delta m^{2}_{21}} \left ( |V^{}_{\tau 1}|^2 - |V^{}_{\tau 2}|^2 \right ) + \frac{m^{2}_{1}}{\Delta m^{2}_{31}} \left ( 1 - 3 |V^{}_{\tau 3}|^2 \right ) \right ]  \; .
\end{equation}

\end{itemize}
\vspace{0.1cm}

We can easily find that the RGEs of $\Phi^{}_{\alpha i}$ and ${\cal J}$ in the cases of IH and ND are alike if neutrinos are Dirac particles. Eqs. (35) and (36) can be obtained from Eqs. (37) and (38) by simply choosing $m^{2}_{1} = - \Delta m^{2}_{31}$. We can expect from above equations (35) - (38) that all nine angles $\Phi^{}_{\alpha i}$ and the Jarlskog ${\cal J}$ may have large evolutions, especially in the MSSM with large $\tan \beta$ if the neutrino mass spectrum is ND or IH.

\subsection{Majorana Neutrinos}

Majorana neutrino masses are believed to be attributed to some physics at a superhigh energy scale $\Lambda$, e.g., the seesaw mechanisms. But all these new physics point to the unique dimension-5 Weinberg operator in an effective theory after the corresponding heavy particles are integrated out \cite{Weinberg:1979sa}
\begin{eqnarray}
{\cal L} & = & \frac{1}{2} \; \overline{l^{}_{L}} H^{}_{} \cdot \kappa \cdot H^{T}_{} l^{c}_{L} \; + \; {\rm h.c.} \; , ~~~~~~ ({\rm SM}) \nonumber \\[3mm]
{\rm or} ~~~~~~ {\cal L} & = & \frac{1}{2} \; \overline{l^{}_{L}} H^{}_{2} \cdot \kappa \cdot H^{T}_{2} l^{c}_{L} \; + \; {\rm h.c.} \; , ~~~~~~ ({\rm MSSM})
\end{eqnarray}
which lead to the effective Majorana neutrino mass matrix $M^{}_{\nu} = \kappa v^2$ (SM) or $M^{}_{\nu} = \kappa v^2 \sin^2\beta$ (MSSM), with $\tan\beta$ denotes the ratio of the vacuum expectation values of two MSSM Higgs doublets. Here $\Lambda$ is a cut off energy scale stands for the energy scale of new physics. The evolution of $\kappa$ from $\Lambda$ down to the electroweak scale $\Lambda^{}_{\rm EW}$ is formally independent of any details of the relevent model from which $\kappa$ is derived. Below $\Lambda$ the energy dependent of the effective neutrino coupling matrix $\kappa$ is described by
\begin{equation}
16\pi^2 \frac{{\rm d}\kappa}{{\rm d}t} \; = \; \alpha^{}_{\rm M} \kappa + C^{}_{} \left [ \left ( Y^{}_{l} Y^{\dagger}_{l} \right ) \kappa + \kappa \left ( Y^{}_{l} Y^{\dagger}_{l} \right )^{T}_{} \right ] \; ,
\end{equation}
at the one-loop level \cite{Majorana RGE}, where $\alpha^{}_{\rm M} \approx - 3 g^{2}_{2} + 6 y^{2}_{t} + \lambda$ (SM) or $\alpha^{}_{\rm M} \approx - 1.2 g^{2}_{1} - 6 g^{2}_{2} + 6 y^{2}_{t}$ (MSSM) with $\lambda$ denotes the Higgs self-coupling in the SM.

Similarly, one may use Eq. (40) to derive the explicit RGEs for neutrino masses and MNS matrix in the flavor basis where $Y^{}_{l}$ is diagonal and real. In this basis, we have $\kappa = V^{}_{} \hat{\kappa} V^{T}_{}$ with $\hat{\kappa} = {\rm diag} \left \{ \kappa^{}_{1}, \; \kappa^{}_{2}, \; \kappa^{}_{3} \right \}$ where $V$ is just the MNS matrix and at $\Lambda^{}_{\rm EW}$ Majorana neutrino masses are $m^{}_{i} = v^{2}_{} \kappa^{}_{i}$ (SM) or $m^{}_{i} = v^{2}_{} \kappa^{}_{i} \sin^{2}_{}\beta$ (MSSM). Then we can further calculate the RGEs of $|V^{}_{\alpha i}|^2$, $\Psi$ ($\Phi$) and ${\cal J}$. We give only the concise results here, where again the excellent approximation of $\tau$-dominance are taken.

\begin{eqnarray}
& & 16\pi^2 \frac{\rm d}{{\rm d}t} \left ( \begin{matrix} |V^{}_{e1}|^2 & |V^{}_{e2}|^2 & |V^{}_{e3}|^2 \cr |V^{}_{\mu 1}|^2 & |V^{}_{\mu 2}|^2 & |V^{}_{\mu 3}|^2 \cr |V^{}_{\tau 1}|^2 & |V^{}_{\tau 2}|^2 & |V^{}_{\tau 3}|^2 \cr \end{matrix} \right ) \nonumber \\ \nonumber \\
& & \; = \; 2 C^{}_{} y^2_\tau \left \{ \frac{m^{2}_{2} + m^{2}_{1}}{\Delta m^{2}_{21}} \left ( \begin{matrix} \; - {^{e1}_{}\Re^{}_{\tau 2}} & ~ ^{e1}_{}\Re^{}_{\tau 2} ~  & ~ 0 ~ \cr \; - {^{\mu1}_{}\Re^{}_{\tau 2}} & ~ ^{\mu1}_{}\Re^{}_{\tau 2} ~ & ~ 0 ~ \cr \; - {^{\tau1}_{}\Re^{}_{\tau 2}} & ~ ^{\tau1}_{}\Re^{}_{\tau 2} ~ & ~ 0 ~ \cr \end{matrix} \right ) 
+ \frac{2 m^{}_{2} m^{}_{1}}{\Delta m^{2}_{21}} \left ( \begin{matrix} \; - {\mathbb R}^{}_{e \tau 1 2} \; & ~ {\mathbb R}^{}_{e \tau 1 2} ~  & ~ 0 ~ \cr \; - {\mathbb R}^{}_{\mu \tau 1 2} \; & ~ {\mathbb R}^{}_{\mu \tau 1 2} ~ & ~ 0 ~ \cr \; - {\mathbb R}^{}_{\tau \tau 1 2} \; & ~ {\mathbb R}^{}_{\tau \tau 1 2} ~ & ~ 0 ~ \cr \end{matrix} \right ) \right. \nonumber \\ \nonumber \\
& & ~~~~~~~~~~~~~~~ \left. + \; \frac{m^{2}_{3} + m^{2}_{1}}{\Delta m^{2}_{31}} \left ( \begin{matrix} \; - {^{e1}_{}\Re^{}_{\tau 3}} & ~ 0 ~ & ^{e1}_{}\Re^{}_{\tau 3} ~  \cr \; - {^{\mu1}_{}\Re^{}_{\tau 3}} & ~ 0 ~ & ^{\mu1}_{}\Re^{}_{\tau 3} ~ \cr \; - {^{\tau1}_{}\Re^{}_{\tau 3}} & ~ 0 ~ & ^{\tau1}_{}\Re^{}_{\tau 3} ~ \cr \end{matrix} \right ) 
+ \frac{2 m^{}_{3} m^{}_{1}}{\Delta m^{2}_{31}} \left ( \begin{matrix} \; - {\mathbb R}^{}_{e \tau 1 3} \; & ~ 0 ~ & ~ {\mathbb R}^{}_{e \tau 1 3} ~  \cr \; - {\mathbb R}^{}_{\mu \tau 1 3} \; & ~ 0 ~ & ~ {\mathbb R}^{}_{\mu \tau 1 3} ~ \cr \; - {\mathbb R}^{}_{\tau \tau 1 3} \; & ~ 0 ~ & ~ {\mathbb R}^{}_{\tau \tau 1 3} ~ \cr \end{matrix} \right ) \right. \nonumber \\ \nonumber \\
& & ~~~~~~~~~~~~~~~ \left. + \; \frac{m^{2}_{3} + m^{2}_{2}}{\Delta m^{2}_{32}} \left ( \begin{matrix} ~ 0 & \; - {^{e2}_{}\Re^{}_{\tau 3}} ~ & ^{e2}_{}\Re^{}_{\tau 3} ~  \cr ~ 0 & \; - {^{\mu2}_{}\Re^{}_{\tau 3}} ~ & ^{\mu2}_{}\Re^{}_{\tau 3} ~ \cr ~ 0 & \; - {^{\tau2}_{}\Re^{}_{\tau 3}} ~ & ^{\tau2}_{}\Re^{}_{\tau 3} ~ \cr \end{matrix} \right ) 
+ \frac{2 m^{}_{3} m^{}_{2}}{\Delta m^{2}_{32}} \left ( \begin{matrix} ~ 0 & \; - {\mathbb R}^{}_{e \tau 2 3} \; & ~ {\mathbb R}^{}_{e \tau 2 3} ~  \cr ~ 0 & \; - {\mathbb R}^{}_{\mu \tau 2 3} \; & ~ {\mathbb R}^{}_{\mu \tau 2 3} ~ \cr ~ 0 & \; - {\mathbb R}^{}_{\tau \tau 2 3} \; & ~ {\mathbb R}^{}_{\tau \tau 2 3} ~ \cr \end{matrix} \right ) \right \} \; , \nonumber \\
\end{eqnarray}

\begin{eqnarray}
& & 16\pi^2 \frac{\rm d}{{\rm d}t} \left ( \begin{matrix} \Psi^{}_{e1} & \Psi^{}_{e2} & \Psi^{}_{e3} \cr \Psi^{}_{\mu 1} & \Psi^{}_{\mu 2} & \Psi^{}_{\mu 3} \cr \Psi^{}_{\tau 1} & \Psi^{}_{\tau 2} & \Psi^{}_{\tau 3} \cr \end{matrix} \right ) \nonumber \\ \nonumber \\
& & \; = \; C^{}_{} y^2_\tau \left \{ \frac{m^{2}_{2} + m^{2}_{1}}{\Delta m^{2}_{21}} \; {\cal J} \left ( \begin{matrix} \; - |V^{}_{e 2}|^{-2} ~ & ~  |V^{}_{e 1}|^{-2} ~ & ~  |V^{}_{e 2}|^{-2} - |V^{}_{e 1}|^{-2} ~ \cr ~ |V^{}_{\mu 2}|^{-2} ~ & \; - |V^{}_{\mu 1}|^{-2} ~ & ~ |V^{}_{\mu 1}|^{-2} - |V^{}_{\mu 2}|^{-2} ~  \cr 0 & 0 & 0 ~  \cr \end{matrix} \right ) \right. \nonumber \\ \nonumber \\
& & ~~~~~~~~~~~~ \left. + \; \frac{m^{2}_{3} + m^{2}_{1}}{\Delta m^{2}_{31}} \; {\cal J} \left ( \begin{matrix} \; - |V^{}_{e 3}|^{-2} ~ & ~ |V^{}_{e 3}|^{-2} - |V^{}_{e 1}|^{-2} ~ & ~ |V^{}_{e 1}|^{-2} ~ \cr \; |V^{}_{\mu 3}|^{-2} ~ & ~ |V^{}_{\mu 1}|^{-2} - |V^{}_{\mu 3}|^{-2} ~ & \; - |V^{}_{\mu 1}|^{-2} ~ \cr 0 & 0 ~ & 0 \cr \end{matrix} \right ) \right. \nonumber \\ \nonumber \\
& & ~~~~~~~~~~~~ \left. + \; \frac{m^{2}_{3} + m^{2}_{2}}{\Delta m^{2}_{32}} \; {\cal J} \left ( \begin{matrix} ~ |V^{}_{e 3}|^{-2} - |V^{}_{e 2}|^{-2} ~ & \; - |V^{}_{e 3}|^{-2} ~ & ~ |V^{}_{e 2}|^{-2} ~ \cr ~ |V^{}_{\mu 2}|^{-2} - |V^{}_{\mu 3}|^{-2} ~ & ~ |V^{}_{\mu 3}|^{-2} ~ & \; - |V^{}_{\mu 2}|^{-2} ~ \cr 0 \; ~ & 0 & 0 \cr \end{matrix} \right ) \right. \nonumber \\ \nonumber \\
& & ~~~~~~~~~~~~ \left. + \; \frac{2 m^{}_{2} m^{}_{1}}{\Delta m^{2}_{21}} \left ( \begin{matrix} ~ {\mathbb I}^{}_{e\tau12} |V^{}_{e 2}|^{-2} ~ & ~ - {\mathbb I}^{}_{e\tau12} |V^{}_{e 1}|^{-2} ~ & ~ {\mathbb I}^{}_{e\tau12} \left ( |V^{}_{e 1}|^{-2} - |V^{}_{e 2}|^{-2} \right ) \; \cr ~ {\mathbb I}^{}_{\mu\tau12} |V^{}_{\mu 2}|^{-2} ~ & ~ - {\mathbb I}^{}_{\mu\tau12} |V^{}_{\mu 1}|^{-2} ~ & ~ {\mathbb I}^{}_{\mu\tau12} \left ( |V^{}_{\mu 1}|^{-2} - |V^{}_{\mu 2}|^{-2} \right ) \; \cr ~ {\mathbb I}^{}_{\tau\tau12} |V^{}_{\tau 2}|^{-2} ~ & ~ - {\mathbb I}^{}_{\tau\tau12} |V^{}_{\tau 1}|^{-2} ~ & ~ {\mathbb I}^{}_{\tau\tau12} \left ( |V^{}_{\tau 1}|^{-2} - |V^{}_{\tau 2}|^{-2} \right ) \; \cr \end{matrix} \right ) \right. \nonumber \\ \nonumber \\
& & ~~~~~~~~~~~~ \left. - \; \frac{2 m^{}_{3} m^{}_{1}}{\Delta m^{2}_{31}} \left ( \begin{matrix} ~ {\mathbb I}^{}_{e\tau13} |V^{}_{e 3}|^{-2} ~ & ~ {\mathbb I}^{}_{e\tau13} \left ( |V^{}_{e 1}|^{-2} - |V^{}_{e 3}|^{-2} \right ) ~ & \; - {\mathbb I}^{}_{e\tau13} |V^{}_{e 1}|^{-2} ~ \cr ~ {\mathbb I}^{}_{\mu\tau13} |V^{}_{\mu 3}|^{-2} ~ & ~ {\mathbb I}^{}_{\mu\tau13} \left ( |V^{}_{\mu 1}|^{-2} - |V^{}_{\mu 3}|^{-2} \right ) ~ & \; - {\mathbb I}^{}_{\mu\tau13} |V^{}_{\mu 1}|^{-2} ~ \cr ~ {\mathbb I}^{}_{\tau\tau13} |V^{}_{\tau 3}|^{-2} ~ & ~ {\mathbb I}^{}_{\tau\tau13} \left ( |V^{}_{\tau 1}|^{-2} - |V^{}_{\tau 3}|^{-2} \right ) ~ & \; - {\mathbb I}^{}_{\tau\tau13} |V^{}_{\tau 1}|^{-2} ~ \cr \end{matrix} \right ) \right. \nonumber \\ \nonumber \\
& & ~~~~~~~~~~~~ \left. + \; \frac{2 m^{}_{3} m^{}_{2}}{\Delta m^{2}_{32}} \left ( \begin{matrix} ~ {\mathbb I}^{}_{e\tau23} \left ( |V^{}_{e 2}|^{-2} - |V^{}_{e 3}|^{-2} \right ) ~ & ~ {\mathbb I}^{}_{e\tau23} |V^{}_{e 3}|^{-2} ~ & ~ - {\mathbb I}^{}_{e\tau23} |V^{}_{e 2}|^{-2} ~ \cr ~ {\mathbb I}^{}_{\mu\tau23} \left ( |V^{}_{\mu 2}|^{-2} - |V^{}_{\mu 3}|^{-2} \right ) ~ & ~ {\mathbb I}^{}_{\mu\tau23} |V^{}_{\mu 3}|^{-2} ~ & ~ - {\mathbb I}^{}_{\mu\tau23} |V^{}_{\mu 2}|^{-2} ~ \cr ~ {\mathbb I}^{}_{\tau\tau23} \left ( |V^{}_{\tau 2}|^{-2} - |V^{}_{\tau 3}|^{-2} \right ) ~ & ~ {\mathbb I}^{}_{\tau\tau23} |V^{}_{\tau 3}|^{-2} ~ & ~ - {\mathbb I}^{}_{\tau\tau23} |V^{}_{\tau 2}|^{-2} ~ \cr \end{matrix} \right ) \right \} \; , \nonumber \\ 
\end{eqnarray}

\begin{eqnarray}
16\pi^2 \frac{\rm d}{{\rm d}t} \; {\cal J} & = & C^{}_{} y^2_\tau \left \{ \frac{m^{2}_{2} + m^{2}_{1}}{\Delta m^{2}_{21}} \; {\cal J} \left ( |V^{}_{\tau 1}|^2 - |V^{}_{\tau 2}|^2 \right ) + \; \frac{m^{2}_{3} + m^{2}_{1}}{\Delta m^{2}_{31}} \; {\cal J} \left ( |V^{}_{\tau 1}|^2 - |V^{}_{\tau 3}|^2 \right ) \right. \nonumber \\[2.4mm]
& & ~ \left. + \; \frac{m^{2}_{3} + m^{2}_{2}}{\Delta m^{2}_{32}} \; {\cal J} \left ( |V^{}_{\tau 2}|^2 - |V^{}_{\tau 3}|^2 \right ) \right. \nonumber \\[2.4mm]
& & ~ \left. + \frac{2 m^{}_{2} m^{}_{1}}{\Delta m^{2}_{21}} \left [ {\mathbb I}^{}_{\tau\tau12} \left ( |V^{}_{\mu 2}|^2 - |V^{}_{\mu 1}|^2 \right ) - {\mathbb I}^{}_{\mu\tau12} \left ( |V^{}_{\tau 2}|^2 - |V^{}_{\tau 1}|^2 \right ) \right ] \right. \nonumber \\[2.4mm]
& & ~ \left. - \frac{2 m^{}_{3} m^{}_{1}}{\Delta m^{2}_{31}} \left [ {\mathbb I}^{}_{\tau\tau13} \left ( |V^{}_{\mu 3}|^2 - |V^{}_{\mu 1}|^2 \right ) - {\mathbb I}^{}_{\mu\tau13} \left ( |V^{}_{\tau 3}|^2 - |V^{}_{\tau 1}|^2 \right ) \right ] \right. \nonumber \\[2.4mm]
& & ~ \left. + \frac{2 m^{}_{3} m^{}_{2}}{\Delta m^{2}_{32}} \left [ {\mathbb I}^{}_{\tau\tau23} \left ( |V^{}_{\mu 3}|^2 - |V^{}_{\mu 2}|^2 \right ) - {\mathbb I}^{}_{\mu\tau23} \left ( |V^{}_{\tau 3}|^2 - |V^{}_{\tau 2}|^2 \right ) \right ] \right \} \; .
\end{eqnarray}

From Eq. (20), we have 
\begin{equation}
16\pi^2 \frac{\rm d}{{\rm d}t} \Phi^{}_{\alpha i} \; = \; 16\pi^2 \left ( \frac{\rm d}{{\rm d}t} \Psi^{}_{\gamma i} - \frac{\rm d}{{\rm d}t} \Psi^{}_{\beta i} \right ) \; ,
\end{equation}
where $\beta$ and $\gamma$ are the next two flavor indices right after $\alpha$. By using this equation, the one-loop RGE of $\Phi$-matrix for the Majorana neutrinos can then be easily obtained from Eq. (42).

We can clearly see that the parameters ${\mathbb R}^{}_{\alpha\beta ij}$ and ${\mathbb I}^{}_{\alpha\beta ij}$ which are associated with the Majorana phases in $V$ and not related to the one-loop RGEs for Dirac neutrinos are involved in the RGEs for Majorana neutrinos. These terms could dominate over others and determine the running behaviours of $|V^{}_{\alpha i}|$, $\Psi^{}_{\alpha i}$ and ${\cal J}$ if the Majorana phases are properly chosen. It is well known that for Dirac neutrinos, if ${\cal J}$ is zero at some scale, it will keep vanished at any energy scale. However, we can see from Eqs. (42) and (43), for Majorana neutrinos, even if ${\cal J} = 0$ (no Dirac type CP violation) at some energy scale, $\Phi^{}_{\alpha i}$ and ${\cal J}$ can still receive significant radiative corrections  only if not all the $\Psi^{}_{\alpha i}$ are zero (i.e., Majorana type CP violation exists).

Here we give the approximate RGEs of $\Psi^{}_{\alpha i}$ and ${\cal J}$ in three limits of neutrino mass hierarchy: NH, IH and ND. 

\begin{itemize}

\item {\bf Normal Hierarchy} ~ $m^{}_{3} \gg m^{}_{2} \gg m^{}_{1} \simeq 0$, $m^{}_{2} \approx \sqrt{\Delta m^{2}_{21}}$ and $m^{}_{3} \approx \sqrt{\Delta m^{2}_{31}}$

In this neutrino masses limit the one-loop RGEs of $\Psi$ and ${\cal J}$ can be approximately expressed as

\begin{eqnarray}
& & 16\pi^2 \frac{\rm d}{{\rm d}t} \left ( \begin{matrix} \Psi^{}_{e1} & \Psi^{}_{e2} & \Psi^{}_{e3} \cr \Psi^{}_{\mu 1} & \Psi^{}_{\mu 2} & \Psi^{}_{\mu 3} \cr \Psi^{}_{\tau 1} & \Psi^{}_{\tau 2} & \Psi^{}_{\tau 3} \cr \end{matrix} \right ) \nonumber\\ \nonumber\\
& & \; \approx \; 2 C^{}_{} y^2_\tau \left \{ {\cal J} \left ( \begin{matrix} - |V^{}_{e 2}|^{-2} & ~ 0 ~ & |V^{}_{e 2}|^{-2} \cr |V^{}_{\mu 2}|^{-2} & ~ 0 ~ & - |V^{}_{\mu 2}|^{-2} \cr 0 & ~ 0 ~ & 0 ~ \cr \end{matrix} \right ) \; + \; \frac{\Delta m^{2}_{21}}{\Delta m^{2}_{31}} {\cal J} \left ( \begin{matrix} ~ |V^{}_{e 3}|^{-2} & - |V^{}_{e 3}|^{-2} & ~ 0 ~ \cr 0 & 0 & 0 \cr 0 & 0 & 0 \cr \end{matrix} \right ) \; \right. \nonumber\\ \nonumber\\
& & ~~~ \left. \; + \; \sqrt{\frac{\Delta m^{2}_{21}}{\Delta m^{2}_{31}}} \left ( \begin{matrix} ~ {\mathbb I}^{}_{e\tau23} \left ( |V^{}_{e 2}|^{-2} - |V^{}_{e 3}|^{-2} \right ) ~ & ~ {\mathbb I}^{}_{e\tau23} |V^{}_{e 3}|^{-2} ~ & ~ - {\mathbb I}^{}_{e\tau23} |V^{}_{e 2}|^{-2} ~ \cr ~ {\mathbb I}^{}_{\mu\tau23} \left ( |V^{}_{\mu 2}|^{-2} - |V^{}_{\mu 3}|^{-2} \right ) ~ & ~ {\mathbb I}^{}_{\mu\tau23} |V^{}_{\mu 3}|^{-2} ~ & ~ - {\mathbb I}^{}_{\mu\tau23} |V^{}_{\mu 2}|^{-2} ~ \cr ~ {\mathbb I}^{}_{\tau\tau23} \left ( |V^{}_{\tau 2}|^{-2} - |V^{}_{\tau 3}|^{-2} \right ) ~ & ~ {\mathbb I}^{}_{\tau\tau23} |V^{}_{\tau 3}|^{-2} ~ & ~ - {\mathbb I}^{}_{\tau\tau23} |V^{}_{\tau 2}|^{-2} ~ \cr \end{matrix} \right ) \right \} \; ,  \nonumber\\
\end{eqnarray}
\begin{eqnarray}
16\pi^2 \frac{\rm d}{{\rm d}t} \; {\cal J} & \approx & 2 C^{}_{} y^2_\tau \; {\cal J} \left ( |V^{}_{\tau 1}|^2 - |V^{}_{\tau 3}|^2 \right ) \nonumber\\
& & + \; 2 C^{}_{} y^2_\tau \; \sqrt{\frac{\Delta m^{2}_{21}}{\Delta m^{2}_{31}}} \left [ {\mathbb I}^{}_{\tau\tau23} \left ( |V^{}_{\mu 3}|^2 - |V^{}_{\mu 2}|^2 \right ) - {\mathbb I}^{}_{\mu\tau23} \left ( |V^{}_{\tau 3}|^2 - |V^{}_{\tau 2}|^2 \right ) \right ] \; . \nonumber\\
\end{eqnarray}
Again, for terms lead by $\Delta m^{2}_{21} / \Delta m^{2}_{31}$, we reserve only those terms inversely proportional to $|V^{}_{e3}|$. By using Eq. (44), the one-loop RGEs of $\Phi^{}_{\alpha i}$ in the limit of NH can be easily derived from Eq. (45):
\begin{eqnarray}
& & 16\pi^2 \frac{\rm d}{{\rm d}t} \left ( \begin{matrix} \Phi^{}_{e1} & \Phi^{}_{e2} & \Phi^{}_{e3} \cr \Phi^{}_{\mu 1} & \Phi^{}_{\mu 2} & \Phi^{}_{\mu 3} \cr \Phi^{}_{\tau 1} & \Phi^{}_{\tau 2} & \Phi^{}_{\tau 3} \cr \end{matrix} \right ) \nonumber\\ \nonumber\\
& & \; \approx \; 2 C^{}_{} y^2_\tau \left \{ \frac{{\cal J}}{|V^{}_{e 2}|^2 |V^{}_{\mu 2}|^2} \left ( \begin{matrix} - |V^{}_{e 2}|^2 & 0 & |V^{}_{e 2}|^2 \cr - |V^{}_{\mu 2}|^2 & 0 & |V^{}_{\mu 2}|^2 \cr \; 1 - |V^{}_{\tau 2}|^2 \; & ~ 0 ~ & \; |V^{}_{\tau 2}|^2 - 1 \; \cr \end{matrix} \right ) \; + \; \frac{\Delta m^{2}_{21}}{\Delta m^{2}_{31}} \; \frac{{\cal J}}{|V^{}_{e 3}|^2} \left ( \begin{matrix} \; 0 & \; 0 & \; 0 ~ \cr \; 1& -1 & \; 0 ~ \cr -1 & \; 1 & \; 0 ~ \cr \end{matrix} \right ) \right. \nonumber\\ \nonumber\\
& & ~~~~~~~~~~~~ \left. \; + \; \sqrt{\frac{\Delta m^{2}_{21}}{\Delta m^{2}_{31}}} \; \frac{\; {\mathbb I}^{}_{e \tau23} \;}{|V^{}_{e 3}|^2} \left ( \begin{matrix} \; \times & \; \times & \; \times ~ \cr -1& \; 1 & \; \times ~ \cr \; 1 & -1 & \; \times ~ \cr \end{matrix} \right ) \right \} \; .
\end{eqnarray}
Here the symbol $\times$ stands for terms that are not inversely proportional to $|V^{}_{e3}|$, which lead only mild corrections to the angle matrix in the case of NH.

Comparing Eqs. (47) and (46) with Eqs. (33) and (34), we can find that the terms lead by $\sqrt{\Delta m^{2}_{21} / \Delta m^{2}_{31}}$ may lead to very different running behaviours of Majorana neutrinos compared to the Dirac neutrinos. For some specific pattern of $V$, terms with ${\mathbb I}^{}_{\alpha\beta ij}$ can be dominating and even change the evolution directions of $\Phi^{}_{\alpha i}$ and ${\cal J}$.

\item {\bf Inverted Hierarchy}  ~ $\sqrt{ - \Delta m^{2}_{31}} \approx m^{}_{2} \approx m^{}_{1} \gg m^{}_{3} \approx 0$

In this neutrino masses limit the one-loop RGEs of $\Phi$ and ${\cal J}$ can be approximately expressed as
\begin{eqnarray}
& & 16\pi^2 \frac{\rm d}{{\rm d}t} \left ( \begin{matrix} \Psi^{}_{e1} & \Psi^{}_{e2} & \Psi^{}_{e3} \cr \Psi^{}_{\mu 1} & \Psi^{}_{\mu 2} & \Psi^{}_{\mu 3} \cr \Psi^{}_{\tau 1} & \Psi^{}_{\tau 2} & \Psi^{}_{\tau 3} \cr \end{matrix} \right ) \nonumber \\ \nonumber \\
& & \; \approx \; - C^{}_{} y^2_\tau \left \{ \frac{4 \Delta m^{2}_{31}}{\Delta m^{2}_{21}} \; {\rm Re} S^{}_{\tau12} \left ( \begin{matrix} ~ \displaystyle \frac{{\rm Im} S^{}_{e12}}{|V^{}_{e 2}|^2} ~ & ~ \displaystyle - \frac{{\rm Im} S^{}_{e12}}{|V^{}_{e 1}|^2} ~ & ~ \displaystyle {\rm Im} S^{}_{e12} \frac{|V^{}_{e 2}|^2 - |V^{}_{e 1}|^2}{|V^{}_{e 1}|^2 |V^{}_{e 2}|^2} ~ \\[3mm] ~ \displaystyle \frac{{\rm Im} S^{}_{\mu12}}{|V^{}_{\mu 2}|^2} ~ & ~ \displaystyle - \frac{{\rm Im} S^{}_{\mu12}}{|V^{}_{\mu 1}|^2} ~ & ~ \displaystyle {\rm Im} S^{}_{\mu12} \frac{|V^{}_{\mu 2}|^2 - |V^{}_{\mu 1}|^2}{|V^{}_{\mu 1}|^2 |V^{}_{\mu 2}|^2} ~ \\[3mm] ~ \displaystyle \frac{{\rm Im} S^{}_{\tau12}}{|V^{}_{\tau 2}|^2} ~ & ~ \displaystyle - \frac{{\rm Im} S^{}_{\tau12}}{|V^{}_{\tau 1}|^2} ~ & ~ \displaystyle{\rm Im} S^{}_{\tau12} \frac{|V^{}_{\tau 2}|^2 - |V^{}_{\tau 1}|^2}{|V^{}_{\tau 1}|^2 |V^{}_{\tau 2}|^2} ~ \\[3mm] \end{matrix} \right ) \right. \nonumber \\ \nonumber \\
& & ~~~~~~~~~~~~~~~~ \left. + \; \left ( \begin{matrix} \; - |V^{}_{e 2}|^{-2} ~ & ~  |V^{}_{e 1}|^{-2} ~ & ~  |V^{}_{e 2}|^{-2} - |V^{}_{e 1}|^{-2} ~ \\[3mm] ~ |V^{}_{\mu 2}|^{-2} ~ & \; - |V^{}_{\mu 1}|^{-2} ~ & ~ |V^{}_{\mu 1}|^{-2} - |V^{}_{\mu 2}|^{-2} ~  \\[3mm] 0 & 0 & 0 ~  \cr \end{matrix} \right ) \right \} \; , \nonumber\\
\end{eqnarray}
\begin{eqnarray}
16\pi^2 \frac{\rm d}{{\rm d}t} \; {\cal J} & = & C^{}_{} y^2_\tau \left \{ \frac{4 \Delta m^{2}_{31}}{\Delta m^{2}_{21}} {\rm Re} S^{}_{\tau12} \left [{\rm Im} S^{}_{\mu12} \left ( |V^{}_{\tau 2}|^2 - |V^{}_{\tau 1}|^2 \right ) - {\rm Im} S^{}_{\tau12} \left ( |V^{}_{\mu 2}|^2 - |V^{}_{\mu 1}|^2 \right ) \right ] \right. \nonumber\\[3mm]
& & \left. ~~~~~~~ + \; {\cal J} \left ( 3 |V^{}_{\tau 3}|^2 - 1 \right ) \right \} \; .
\end{eqnarray}
Note that, in the case of IH, $\Delta m^{2}_{31}$ is negative.

\item {\bf Near Degeneracy}  ~ $m^{}_{3} \approx m^{}_{2} \approx m^{}_{1} $ and $\Delta m^{2}_{32} \approx \Delta m^{2}_{31}$.

If three neutrino masses are nearly degenerate the one-loop RGEs of $\Psi$ and ${\cal J}$ can be approximately expressed as

\begin{eqnarray}
& & 16\pi^2 \frac{\rm d}{{\rm d}t} \left ( \begin{matrix} \Psi^{}_{e1} & \Psi^{}_{e2} & \Psi^{}_{e3} \cr \Psi^{}_{\mu 1} & \Psi^{}_{\mu 2} & \Psi^{}_{\mu 3} \cr \Psi^{}_{\tau 1} & \Psi^{}_{\tau 2} & \Psi^{}_{\tau 3} \cr \end{matrix} \right ) \nonumber \\ \nonumber \\
& & \; \approx \; 4 C^{}_{} y^2_\tau \left \{ \frac{m^{2}_{1}}{\Delta m^{2}_{21}} \; {\rm Re} S^{}_{\tau12} \left ( \begin{matrix} ~ \displaystyle \frac{{\rm Im} S^{}_{e12}}{|V^{}_{e 2}|^2} ~ & ~ \displaystyle - \frac{{\rm Im} S^{}_{e12}}{|V^{}_{e 1}|^2} ~ & ~ \displaystyle {\rm Im} S^{}_{e12} \frac{|V^{}_{e 2}|^2 - |V^{}_{e 1}|^2}{|V^{}_{e 1}|^2 |V^{}_{e 2}|^2} ~ \\[3mm] ~ \displaystyle \frac{{\rm Im} S^{}_{\mu12}}{|V^{}_{\mu 2}|^2} ~ & ~ \displaystyle - \frac{{\rm Im} S^{}_{\mu12}}{|V^{}_{\mu 1}|^2} ~ & ~ \displaystyle {\rm Im} S^{}_{\mu12} \frac{|V^{}_{\mu 2}|^2 - |V^{}_{\mu 1}|^2}{|V^{}_{\mu 1}|^2 |V^{}_{\mu 2}|^2} ~ \\[3mm] ~ \displaystyle \frac{{\rm Im} S^{}_{\tau12}}{|V^{}_{\tau 2}|^2} ~ & ~ \displaystyle - \frac{{\rm Im} S^{}_{\tau12}}{|V^{}_{\tau 1}|^2} ~ & ~ \displaystyle{\rm Im} S^{}_{\tau12} \frac{|V^{}_{\tau 2}|^2 - |V^{}_{\tau 1}|^2}{|V^{}_{\tau 1}|^2 |V^{}_{\tau 2}|^2} ~ \\[3mm] \end{matrix} \right ) \right. \nonumber \\ \nonumber \\
& & ~~~~~~~~~~~ \left. + \; \frac{m^{2}_{1}}{\Delta m^{2}_{31}} \left [ - {\rm Re} S^{}_{\tau13} \left ( \begin{matrix} ~ \displaystyle \frac{{\rm Im} S^{}_{e13}}{|V^{}_{e 3}|^2} ~ & ~ \displaystyle {\rm Im} S^{}_{e13} \frac{|V^{}_{e 3}|^2 - |V^{}_{e 1}|^2}{|V^{}_{e 1}|^2 |V^{}_{e 3}|^2} ~ & ~ \displaystyle - \frac{{\rm Im} S^{}_{e13}}{|V^{}_{e 1}|^2} ~ \\[3mm] ~ \displaystyle \frac{{\rm Im} S^{}_{\mu13}}{|V^{}_{\mu 3}|^2} ~ & ~ \displaystyle {\rm Im} S^{}_{\mu13} \frac{|V^{}_{\mu 3}|^2 - |V^{}_{\mu 1}|^2}{|V^{}_{\mu 1}|^2 |V^{}_{\mu 3}|^2} ~ & ~ \displaystyle - \frac{{\rm Im} S^{}_{\mu13}}{|V^{}_{\mu 1}|^2} ~ \\[3mm] ~ \displaystyle \frac{{\rm Im} S^{}_{\tau13}}{|V^{}_{\tau 3}|^2} ~ & ~ \displaystyle{\rm Im} S^{}_{\tau13} \frac{|V^{}_{\tau 3}|^2 - |V^{}_{\tau 1}|^2}{|V^{}_{\tau 1}|^2 |V^{}_{\tau 3}|^2} ~ & ~ \displaystyle - \frac{{\rm Im} S^{}_{\tau13}}{|V^{}_{\tau 1}|^2} ~ \\[3mm] \end{matrix} \right ) \right. \right. \nonumber \\ \nonumber \\
& & ~~~~~~~~~~~~~~~~~~~~~~ \left. \left. + \; {\rm Re} S^{}_{\tau23} \left ( \begin{matrix} ~ \displaystyle {\rm Im} S^{}_{e23} \frac{|V^{}_{e 3}|^2 - |V^{}_{e 2}|^2}{|V^{}_{e 2}|^2 |V^{}_{e 3}|^2} ~ & ~ \displaystyle \frac{{\rm Im} S^{}_{e23}}{|V^{}_{e 3}|^2} ~ & ~ \displaystyle - \frac{{\rm Im} S^{}_{e23}}{|V^{}_{e 2}|^2} ~ \\[3mm] ~ \displaystyle {\rm Im} S^{}_{\mu23} \frac{|V^{}_{\mu 3}|^2 - |V^{}_{\mu 2}|^2}{|V^{}_{\mu 2}|^2 |V^{}_{\mu 3}|^2} ~ & ~ \displaystyle \frac{{\rm Im} S^{}_{\mu23}}{|V^{}_{\mu 3}|^2} ~ & ~ \displaystyle - \frac{{\rm Im} S^{}_{\mu23}}{|V^{}_{\mu 2}|^2} ~ \\[3mm] ~ \displaystyle{\rm Im} S^{}_{\tau23} \frac{|V^{}_{\tau 3}|^2 - |V^{}_{\tau 2}|^2}{|V^{}_{\tau 2}|^2 |V^{}_{\tau 3}|^2} ~ & ~ \displaystyle \frac{{\rm Im} S^{}_{\tau23}}{|V^{}_{\tau 3}|^2} ~ & ~ \displaystyle - \frac{{\rm Im} S^{}_{\tau23}}{|V^{}_{\tau 2}|^2} ~ \\[3mm] \end{matrix} \right ) \right ] \right \} \; , \nonumber\\
\end{eqnarray}
\begin{eqnarray}
16\pi^2 \frac{\rm d}{{\rm d}t} \; {\cal J} & = & 4 C^{}_{} y^2_\tau \left \{ \frac{m^{2}_{1}}{\Delta m^{2}_{21}} {\rm Re} S^{}_{\tau12} \left [ {\rm Im} S^{}_{\tau12} \left ( |V^{}_{\mu 2}|^2 - |V^{}_{\mu 1}|^2 \right ) - {\rm Im} S^{}_{\mu12} \left ( |V^{}_{\tau 2}|^2 - |V^{}_{\tau 1}|^2 \right ) \right ] \right. \nonumber\\
& & \left. \; - \; \frac{m^{2}_{1}}{\Delta m^{2}_{31}} \; {\rm Re} S^{}_{\tau13} \left [ {\rm Im} S^{}_{\tau13} \left ( |V^{}_{\mu 3}|^2 - |V^{}_{\mu 1}|^2 \right ) - {\rm Im} S^{}_{\mu13} \left ( |V^{}_{\tau 3}|^2 - |V^{}_{\tau 1}|^2 \right ) \right ] \right. \nonumber\\
& & \left. \; + \; \frac{m^{2}_{1}}{\Delta m^{2}_{31}} \; {\rm Re} S^{}_{\tau23} \left [ {\rm Im} S^{}_{\tau23} \left ( |V^{}_{\mu 3}|^2 - |V^{}_{\mu 2}|^2 \right ) - {\rm Im} S^{}_{\mu23} \left ( |V^{}_{\tau 3}|^2 - |V^{}_{\tau 2}|^2 \right ) \right ] \right \} \; . \nonumber\\
\end{eqnarray}

\end{itemize}

Eqs. (45) - (51) indicate that the RGE running behaviours of $\Phi$, $\Psi$ and ${\cal J}$ for Majorana neutrinos are very different to that for Dirac neutrinos (see Eqs. (33) - (38)). We can find that three neutrino masses are nearly degenerate, the running behaviours of $\Phi$, $\Psi$ and ${\cal J}$ depend on the interplay of several terms (see Eqs. (37), (38), (50) and (51)) and are very sensitive to the sign of $\Delta m^{2}_{31}$ no matter whether the neutrinos are Dirac or Majorana particles.

\section{Numerical Analysis and Discussion}

The running behaviours of the above mentioned rephasing-invariant quantities are numerically illustrated by assuming $\Lambda \sim 10^{14}$ GeV, which is the typical scale of the conventional seesaw mechanisms and is very close to the scale of the grand unified theories. We chose several sets of typical values of the angle matrix elements in Eq. (9) (Eq. (16)) which are allowed by current $3 \sigma$ experimental data at $\Lambda^{}_{\rm EW}$ and calculated the RGE running effects of the Dirac (Majorana) angle matrix $\Phi$ ($\Psi$) and the Jarlskog invariant ${\cal J}$ from $\Lambda^{}_{\rm EW}$ up to $\Lambda$. For each set of inputs, we consider four typical pattterns of neutrino mass spectrum: i) NH ($m^{}_{1} \simeq 0$) , ii) IH ($m^{}_{3} \simeq 0$) , iii) ND with $\Delta m^{2}_{31} > 0$ and iv) ND with $\Delta m^{2}_{31} < 0$. In our numerical calculation, $\Delta m^{2}_{21} = 7.59 \times 10^{-5} ~ {\rm eV}^2$ and $\Delta m^{2}_{31} = \pm 2.4 \times 10^{-3} ~ {\rm eV}^2$ have been taken as the typical inputs at $\Lambda^{}_{\rm EW}$ and in case iii) and iv) we have chosen $m^{}_{1} = 0.2$ eV. We carry out our numerical calculation in the framework of either the SM or the MSSM, where the Higgs mass $m^{}_{H} = 140$ GeV in the SM and the parameter $\tan\beta = 10$ or 50 in the MSSM have typically been input. Our numerical results and the corresponding inputs are summarized in Tables I - IV and Figs. 2 - 5.

In Table I-III, for the same set of inputs, we calculated the radiative corrections to ${\cal J}$, $\Phi$ and $\Psi$ at $\Lambda$ in the SM (Table I), the MSSM with $\tan\beta = 10$ (Table II) and the MSSM with $\tan\beta = 50$ (Table III) respectively. If the neutrino mass spectrum is NH or IH, angles of $\Phi$ and $\Psi$ can receive non-negligible radiative corrections (larger than $1^{\circ}_{}$) only in the case of MSSM with $\tan\beta = 50$. If three neutrino masses are nearly degenerate, all the angles of $\Phi$ and $\Psi$ can receive significant radiative corrections especially in the MSSM and we can find that their running behaviours are very sensitive to the sign of $\Delta m^{2}_{31}$ for either Majorana or Dirac neutrinos.

The running behaviours of the Jarlskog invariant ${\cal J}$ in the SM and the MSSM with $\tan\beta = 10$ are shown in Fig. 2 and 3 respectively for both the Dirac and Majorana neutrinos. The Jarlskog ${\cal J}$ can receive significant radiative correction if three neutrino masses are nearly degenerate. As already mentioned in Section III, if neutrinos are Majorana particles, the running of ${\cal J}$ is very sensitive to the sign of $\Delta m^{2}_{31}$ in case of ND. For this  specific set of inputs (with negative ${\cal J}$), we can see that if $\Delta m^{2}_{31} < 0$, ${\cal J}$ will decrease when running from $\Lambda^{}_{\rm EW}$ to $\Lambda$. If $\Delta m^{2}_{31} > 0$, ${\cal J}$ will increase and in case of MSSM with $\tan\beta = 50$, ${\cal J}$ can even run above zero and evolve to a positive value (see Case (a) in Fig. 5).

In the framework of MSSM with $\tan\beta = 50$, we consider another two sets of inputs with ${\cal J} > 0$ and ${\cal J} = 0$ respectively. The corresponding results are shown in Table IV and V. Fig. 4 and 5 illustrate the evolution of ${\cal J}$ in these three special cases: a) ${\cal J} < 0$ (Table III), b) ${\cal J} > 0$ (Table IV) and c) ${\cal J} = 0$ (Table V) for Dirac and Majorana neutrinos respectively. Note that the input values of ${\cal J}$ and $\Phi^{}_{\alpha i}$ at $\Lambda^{}_{\rm EW}$ in Table IV have the same absolute values but opposite signs as that in Table III. If neutrinos are Dirac particles, the RGE running behaviours of ${\cal J}$ and $\Phi^{}_{\alpha i}$ in these two cases are entirely opposite. We can find that at $\Lambda \sim 10^{14}$ GeV, ${\cal J}$ and $\Phi^{}_{\alpha i}$ in Table III and IV still have the same absolute values but opposite signs. For Dirac neutrinos, $\left | {\cal J} \right |$ is always decreasing and approaching zero in the MSSM when running from $\Lambda^{}_{\rm EW}$ to $\Lambda$. It means that when evolving from $\Lambda^{}_{\rm EW}$ to $\Lambda$, positive ${\cal J}$ decreases while negative ${\cal J}$ increases on the contrary, but the signs of ${\cal J}$ are not changed. However, if neutrinos are Majorana particles, the running behaviours of the Dirac-type CP-violating parameters ${\cal J}$ and $\Phi^{}_{\alpha i}$ strongly depend on the Majorana-type CP-violating parameters $\Psi^{}_{\alpha i}$. A point need to be pointed out is that the running behaviours of ${\cal J}$ in a), b) and c) three cases are somehow similar as shown in Fig. 5. This is because of these three sets of inputs correspond to the same values of the Majorana angles $\rho$ and $\sigma$ in the Standard Parametrization. By choosing a different set of inputs of $\Psi^{}_{\alpha i}$, we are able to make a very different evolution of ${\cal J}$ for the same set of $\Phi^{}_{\alpha i}$. A more special case is that ${\cal J} = 0$ but the Majorana CP violation is nonzero as shown in Table V. If ${\cal J} = 0$ is input at $\Lambda^{}_{\rm EW}$, ${\cal J}$ will keep vanishing in case of Dirac neutrinos. But for Majorana neutrinos, ${\cal J}$ can evolve to a nonzero value at $\Lambda$, which indicates that the UTs expand from a line and the Dirac-type CP violation is radiatively generated from the Majorana-type CP violations.

As we have declaimed in  the beginning section, both the Dirac-type and the Majorana-type CP violation can be illustrated by the six UTs. Therefore, the RGE running of the Dirac angle matrix $\Phi$ and the Majorana angle matrix $\Psi$ also correspond to the evolutions and the rotations of the UTs in the complex plane. We choose two examples in the MSSM with $\tan\beta = 50$ where the Majorana neutrino masses are nearly degenerate and show in Fig. 6 ($\Delta m^{2}_{31} > 0$) and 7 ($\Delta m^{2}_{31} < 0$) how the six UTs evolve from $\Lambda^{}_{\rm EW}$ to $\Lambda$, where triangles with thicker sides are at higher energy scale. The common area of the six UTs equals to ${\cal J} / 2$. Since the orientations of $\triangle^{}_{e}$, $\triangle^{}_{\mu}$ and $\triangle^{}_{\tau}$ have no physical meaning, we simply choose one side of each triangle ($V^{}_{\mu 3} V^{*}_{\tau 3}$, $V^{}_{\tau 3} V^{*}_{e 3}$ and $V^{}_{e 3} V^{*}_{\mu 3}$) to lie on the x-axial and point to the origin. In this way, the triangle lie above the x-axial is clockwise and corresponds to a positive ${\cal J}$ and the triangle lie below x-axial is anti-clockwise corresponds to a negative ${\cal J}$ on the contrary. Then we can clearly see from Fig. 6, when evolve from $\Lambda^{}_{\rm EW}$ to the high energy scale $\Lambda$, the Jarlskog ${\cal J}$ changed its sign as illustrated in Table III.

\section{Summary}

In summary, we introduced the concepts of the Dirac angle matrix $\Phi$ and the Majorana angle matrix $\Psi$ for Dirac and Majorana neutrinos respectively and show that the angle matrix carries equivalent information to the complex mixing matrix itself, but with the added advantage of being basis and phase convention independent. Our prescription works for any number of fermion generation. We further calculated the one-loop RGEs of $\Phi$, $\Psi$ and some other rephasing invariant parameters. Numerical analyses are carried out for illustration. We find that apparently different from the case of Dirac neutrinos, for Majorana neutrinos the RG-evolutions of  $\Phi$, $\Psi$ and ${\cal J}$ strongly depend on the Majorana-type CP-violating parameters and are quite sensitive to the sign of $\Delta m^{2}_{31}$. They may receive significant radiative corrections in the MSSM with large $\tan\beta$ if three neutrino masses are nearly degenerate.

Of course, the numerical examples presented in this work are mainly for the purpose of illustration. The point is that the nature and the mass spectrum of neutrinos determine the RGE running behaviours of those rephasing invariant parameters which may be crucial for building a realistic neutrino model. Our analysis complement those previous studies of radiative corrections to the physical parameters of Dirac and Majorana neutrinos and are helpful for building a realistic neutrino mass model at a high energy scale.

\begin{acknowledgements}

I would like to thank Prof. Z. Z. Xing for enlightening me on this subject. This work was supported in part by the National Basic Research Program (973 Program) of China under Grant No. 2009CB824800, the National Natural Science Foundation of China under Grant No. 11105113, the Fujian Provincial Natural Science Foundation under Grant No. 2011J05012 and the China Postdoctoral Science Foundation funded project under Grant No. 201104340.

\end{acknowledgements}

\begin{appendix}

\section{Reconstruct the Leptonic Mixing Matrix $V$ from the Dirac Angle Matrix $\Phi$ or the Majorana Angle Matrix $\Psi$}

In Section II we have shown that the Dirac (Majorana) angle matrix carries equivalent independent real parameters as that of the MNS matrix $V$ for Dirac (Majorana) neutrinos. We now further prove the equivalence of the mixing matrix and the angle matrix in the presence of the Dirac-type CP violation (which indicates that none of the moduli $|V^{}_{\alpha i}|$ is zero) by presenting the way of re-obtaining the mixing matrix starting from the angle matrix in case of three generation neutrinos. We shall take it as given that the matrix of mixing moduli is essentially equivalent to the complex mixing matrix \cite{reconstruction}, and content ourselves in the first instance with showing how to obtain the mixing-matrix moduli $|V^{}_{\alpha i}|$, starting from the angles.

\subsection{Dirac Angle Matrix $\Phi$}

The $3 \times 3$ Dirac angle matrix $\Phi$ is defined by Eqs. (9) and (10), and its nine matrix elements satisfy the normalization conditions of Eq. (11). The Dirac angle matrix can be manipulated to yield the magnitudes of mixing matrix elements. Firstly, we introduce the $\sin$-matrix 
\begin{equation}
\sin\Phi \; = \; \left ( \begin{matrix} \sin\Phi^{}_{e1} & ~\sin\Phi^{}_{e2}~ & \sin\Phi^{}_{e3} \cr \sin\Phi^{}_{\mu 1} & \sin\Phi^{}_{\mu 2} & \sin\Phi^{}_{\mu 3} \cr \sin\Phi^{}_{\tau 1} & \sin\Phi^{}_{\tau 2} & \sin\Phi^{}_{\tau 3} \end{matrix} \right ) \; = \; {\cal J} \left ( \begin{matrix} |^{\mu 2}_{}\Box^{}_{\tau 3}|^{-1} & ~|^{\mu 3}_{}\Box^{}_{\tau 1}|^{-1}~ & |^{\mu 1}_{}\Box^{}_{\tau 2}|^{-1} \cr |^{\tau 2}_{}\Box^{}_{e 3}|^{-1} & |^{\tau 3}_{}\Box^{}_{e 1}|^{-1} & |^{\tau 1}_{}\Box^{}_{e 2}|^{-1} \cr |^{e 2}_{}\Box^{}_{\mu 3}|^{-1} & |^{e 3}_{}\Box^{}_{\mu 1}|^{-1} & |^{e 1}_{}\Box^{}_{\mu 2}|^{-1} \end{matrix} \right ) \; ,
\end{equation}
and then define certain products of sines $\Xi^{}_{\alpha i}$:
\begin{equation}
\Xi^{}_{\alpha i} \; \equiv \; \sin\Phi^{}_{\alpha j} \sin\Phi^{}_{\alpha k} \sin\Phi^{}_{\beta i} \sin\Phi^{}_{\gamma i} \; ,
\end{equation}
multiplying together the four $\sin\Phi$ entries in the same row and column as $\sin\Phi^{}_{\alpha i}$, excluding $\sin\Phi^{}_{\alpha i}$ itself. Clearly every mixing modulus-squared except $|V^{}_{\alpha i}|^2$ enters in the denominator of the product $\Xi^{}_{\alpha i}$,whereby the $\Xi^{}_{\alpha i}$ must be proportional to $|V^{}_{\alpha i}|^2$,
\begin{equation}
\Xi^{}_{\alpha i} \; = \; \frac{1}{N} \; |V^{}_{\alpha i}|^2 \; .
\end{equation}
 The relevent normalising factor $N$ may be obtained by summing over any row or column (or indeed over both rows and columns)
\begin{equation}
\frac{1}{N} \; = \; \sum^{}_{\alpha} \Xi^{}_{\alpha i} \; = \; \sum^{}_{i} \Xi^{}_{\alpha i} \; = \; \frac{1}{3} \sum^{}_{\alpha, i} \Xi^{}_{\alpha i} \; = \;  \frac{\displaystyle {\cal J}^4}{ \; \displaystyle \prod^{}_{\alpha, i} |V^{}_{\alpha i}|^2} \; .
\end{equation}
Then we have
\begin{equation}
|V^{}_{\alpha i}|^2 \; = \; N \; \Xi^{}_{\alpha i} \; = \; \frac{\displaystyle 3 \; \Xi^{}_{\alpha i}}{\displaystyle \sum^{}_{\beta j} \Xi^{}_{\beta j}} \; .
\end{equation}
As for the Jarlskog ${\cal J}$, we can find from Eq. (A1) that
\begin{equation}
\prod^{}_{\alpha, i} \sin\Phi^{}_{\alpha i} \; = \; \frac{\displaystyle {\cal J}^9}{\left ( \displaystyle \prod^{}_{\beta, j} |V^{}_{\beta j}|^2 \right )^2} \; ,
\end{equation}
together with Eq. (A4) and (A5), we can obtain 
\begin{equation}
{\cal J} \; = \; \frac{\displaystyle 9 \prod^{}_{\alpha, i} \sin\Phi^{}_{\alpha i}}{\left ( \displaystyle \sum^{}_{\alpha, i} \Xi^{}_{\alpha i} \right )^2} \; .
\end{equation}
Then we conclude that the $\Phi$-matrix is equivalent to the complex mixing matrix $V$ if neutrinos are Dirac particles.

\subsection{Majorana Angle Matrix $\Psi$}

The Majorana angle matrix is defined as
\begin{equation}
\Psi \; = \; \left ( \begin{matrix} \Psi^{}_{e1} & ~\Psi^{}_{e2}~ & \Psi^{}_{e3} \cr \Psi^{}_{\mu 1} & \Psi^{}_{\mu 2} & \Psi^{}_{\mu 3} \cr \Psi^{}_{\tau 1} & \Psi^{}_{\tau 2} & \Psi^{}_{\tau 3} \end{matrix} \right ) \; = \; \left ( \begin{matrix} \arg \left ( V^{}_{e 2} V^{*}_{e 3} \right ) & ~\arg \left ( V^{}_{e 3} V^{*}_{e 1} \right )~ & \arg \left ( V^{}_{e 1} V^{*}_{e 2} \right ) \cr \arg \left ( V^{}_{\mu 2} V^{*}_{\mu 3} \right ) & \arg \left ( V^{}_{\mu 3} V^{*}_{\mu 1} \right ) & \arg \left ( V^{}_{\mu 1} V^{*}_{\mu 2} \right ) \cr \arg \left ( V^{}_{\tau 2} V^{*}_{\tau 3} \right ) & \arg \left ( V^{}_{\tau 3} V^{*}_{\tau 1} \right ) & \arg \left ( V^{}_{\tau 1} V^{*}_{\tau 2} \right ) \end{matrix} \right ) \; ,
\end{equation}
from which we can easily obtain the $\Phi$-matrix defined in Eqs. (9) and (10) by using the relation of Eq. (20). Follow the same procedure as that in above section, we can obtain the $|V|$-matrix and the Jarlskog ${\cal J}$ from the $\Phi$-matrix. Two extra Majorana phases can also be deduced from the $\Psi$-matrix itself.

For a specific parametrization of $V$, for example the Standard Parametrization
\begin{equation}
V \; = \; \left ( \begin{matrix} e^{i \alpha}_{} & & \cr & e^{i \beta}_{} & \cr & & e^{i \gamma}_{} \end{matrix} \right ) \left ( \begin{matrix} c^{}_{12} c^{}_{13} & s^{}_{12} c^{}_{13} & s^{}_{13} e^{- i \delta}_{} \cr -s^{}_{12} c^{}_{23} - c^{}_{12} s^{}_{23} s^{}_{13} e^{i \delta}_{} & c^{}_{12} c^{}_{23} - s^{}_{12} s^{}_{23} s^{}_{13} e^{i \delta}_{} & s^{}_{23} c^{}_{13} \cr s^{}_{12} s^{}_{23} - c^{}_{12} c^{}_{23} s^{}_{13} e^{i \delta}_{} & - c^{}_{12} s^{}_{23} - s^{}_{12} c^{}_{23} s^{}_{13} e^{i \delta}_{} & c^{}_{23} c^{}_{13} \end{matrix} \right ) \left ( \begin{matrix} e^{i \rho}_{} & & \cr & e^{i \sigma}_{} & \cr & & 1 \end{matrix} \right ) \; ,
\end{equation}
three mixing angles and the Dirac phase $\delta$ can be determined from the $|V|$-matrix and the Jarlskog ${\cal J}$:
\begin{eqnarray}
\sin\theta^{}_{13} & = & |V^{}_{e 3}| \; , \\
\tan\theta^{}_{12} & = & |V^{}_{e 2}| / |V^{}_{e 1}| \; , \\
\tan\theta^{}_{23} & = & |V^{}_{\mu 3}| / |V^{}_{\tau 3}| \; ,
\end{eqnarray}
\begin{eqnarray}
\sin\delta & = & {\cal J} \left ( 1-|V^{}_{e 3}|^2 \right ) / |V^{}_{e 1}| |V^{}_{e 2}| |V^{}_{e 3}| |V^{}_{\mu 3}| |V^{}_{\tau 3}| \; , \\[2.4mm]
\cos\delta & = & \frac{\left ( |V^{}_{e 1}|^2 |V^{}_{\mu 1}|^2 - |V^{}_{e 2}|^2 |V^{}_{\mu 2}|^2 \right ) |V^{}_{\tau 3}|^2 - \left ( |V^{}_{e 1}|^2 |V^{}_{\tau 1}|^2 - |V^{}_{e 2}|^2 |V^{}_{\tau 2}|^2 \right ) |V^{}_{\mu 3}|^2}{|V^{}_{e 1}| |V^{}_{e 2}| |V^{}_{e 3}| |V^{}_{\mu 3}| |V^{}_{\tau 3}|} \; .
\end{eqnarray}
Both the absolute value and the quadrant  of $\delta$ can be determined by Eqs. (A13) and (A14). Two Majorana phases $\rho$ and $\sigma$ can also be easily obtained through
\begin{eqnarray}
\rho & = & \delta - \Psi^{}_{e 2} \; , \\
\sigma & = & \Psi^{}_{e 1} - \delta \; .
\end{eqnarray}

\end{appendix}

\newpage

\begin{figure}
\begin{center}
\vspace{14cm}
\includegraphics[bbllx=6.5cm, bblly=6.0cm, bburx=15.0cm, bbury=14.2cm, width=7.9cm, height=7.9cm, angle=0, clip=0]{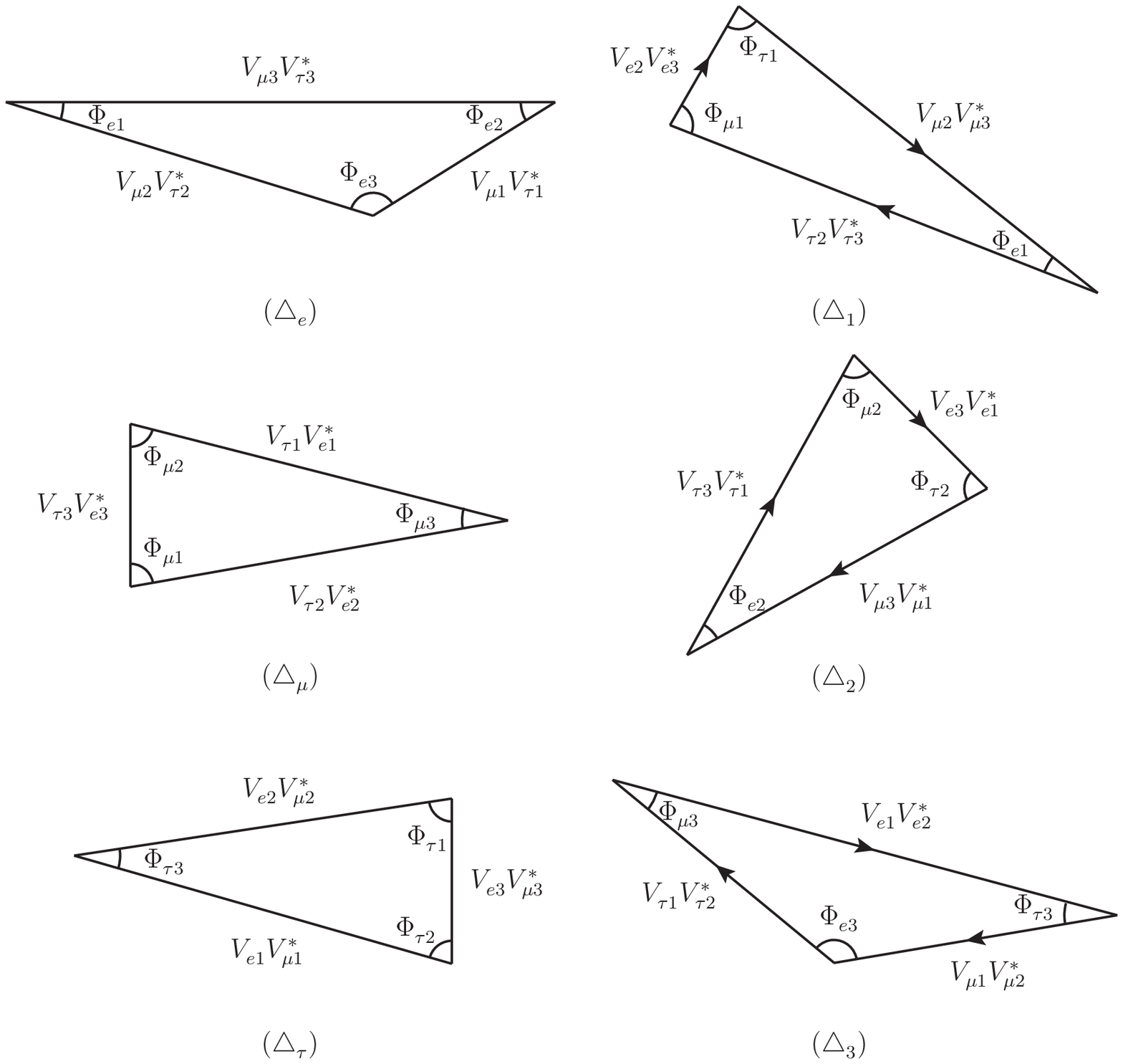}
\vspace{-3cm}\caption{Schematic diagrams for six leptonic unitarity triangles in the complex plane, where each triangle is named by the index that does not manifest in its three sides.} 
\end{center}
\end{figure}

\begin{table}[h]
\caption{Radiative corrections to the Jarlskog ${\cal J}$, the Dirac angle matrix $\Phi$ and the Majorana angle matrix $\Psi$ at $\Lambda \sim 10^{14}$ GeV in the SM (with $m^{}_{H} = 140$ GeV) where we choose $\theta^{}_{12} = 34^{\circ}$, $\theta^{}_{23} = 46^{\circ}$, $\theta^{}_{13} = 7^{\circ}$, $\delta = -90^{\circ}$, $\rho = 120^{\circ}$ and $\sigma = 60^{\circ}$ (in the standard parametrization of $V$) as typical inputs at the electroweak energy scale $\Lambda^{}_{\rm EW}$.}
\begin{center}
\begin{tabular}{c|c|c|c|c|c|c|c|c|c}
\hline
\hline
& $\Lambda^{}_{\rm EW} $ & \multicolumn{8}{c}{$\Lambda \sim 10^{14}$ GeV} \\
\cline{3-10}
& & \multicolumn{4}{c|}{Majorana Neutrinos} & \multicolumn{4}{c}{Dirac Neutrinos} \\
\cline{3-10}
& & NH & IH &  \multicolumn{2}{c|}{Near Degeneracy} & NH & IH &  \multicolumn{2}{c}{Near Degeneracy} \\
\cline{5-6} \cline{9-10}
& & $m^{}_{1} \simeq 0$ & $m^{}_{3} \simeq 0$ & $\Delta m^{2}_{31} > 0$ & $\Delta m^{2}_{31} < 0$ & $m^{}_{1} \simeq 0$ & $m^{}_{3} \simeq 0$ & $\Delta m^{2}_{31} > 0$ & $\Delta m^{2}_{31} < 0$ \\
\hline
\hline
${\cal J}$ & -0.027812 & -0.027813 & -0.027815 & -0.027986 & -0.027748 & -0.027813 & -0.027823 & -0.028010 & -0.027993 \\  
\hline
\hline
$\Phi^{}_{e1}$ & $-9.40^{\circ}$ & $-9.40^{\circ}$ & $-9.41^{\circ}$ & $-9.50^{\circ}$ & $-9.41^{\circ}$ & $-9.40^{\circ}$ & $-9.42^{\circ}$ & $-9.58^{\circ}$ & $-9.58^{\circ}$ \\  
\hline
$\Phi^{}_{e2}$ & $-20.50^{\circ}$ & $-20.50^{\circ}$ & $-20.49^{\circ}$ & $-20.48^{\circ}$ & $-20.30^{\circ}$ & $-20.50^{\circ}$ & $-20.47^{\circ}$ & $-20.15^{\circ}$ & $-20.10^{\circ}$ \\
\hline
$\Phi^{}_{e3}$ & $-150.10^{\circ}$ & $-150.10^{\circ}$ & $-150.10^{\circ}$ & $-150.02^{\circ}$ & $-150.29^{\circ}$ & $-150.10^{\circ}$ & $-150.11^{\circ}$ & $-150.27^{\circ}$ & $-150.32^{\circ}$ \\ 
\hline
$\Phi^{}_{\mu1}$ & $-85.46^{\circ}$ & $-85.46^{\circ}$ & $-85.43^{\circ}$ & $-84.93^{\circ}$ & $-84.91^{\circ}$ & $-85.46^{\circ}$ & $-85.47^{\circ}$ & $-85.64^{\circ}$ & $-85.64^{\circ}$ \\
\hline
$\Phi^{}_{\mu2}$ & $-80.10^{\circ}$ & $-80.10^{\circ}$ & $-80.13^{\circ}$ & $-80.59^{\circ}$ & $-80.72^{\circ}$ & $-80.10^{\circ}$ & $-80.10^{\circ}$ & $-80.03^{\circ}$ & $-80.02^{\circ}$ \\
\hline
$\Phi^{}_{\mu3}$ & $-14.44^{\circ}$ & $-14.44^{\circ}$ & $-14.44^{\circ}$ & $-14.48^{\circ}$ & $-14.37^{\circ}$ & $-14.44^{\circ}$ & $-14.43^{\circ}$ & $-14.33^{\circ}$ & $-14.34^{\circ}$ \\
\hline
$\Phi^{}_{\tau1}$ & $-85.14^{\circ}$ & $-85.14^{\circ}$ & $-85.16^{\circ}$ & $-85.57^{\circ}$ & $-85.68^{\circ}$ & $-85.14^{\circ}$ & $-85.11^{\circ}$ & $-84.78^{\circ}$ & $-84.78^{\circ}$ \\ 
\hline
$\Phi^{}_{\tau2}$ & $-79.40^{\circ}$ & $-79.40^{\circ}$ & $-79.38^{\circ}$ & $-78.93^{\circ}$ & $-78.98^{\circ}$ & $-79.40^{\circ}$ & $-79.43^{\circ}$ & $-79.82^{\circ}$ & $-79.88^{\circ}$ \\
\hline
$\Phi^{}_{\tau3}$ & $-15.46^{\circ}$ & $-15.46^{\circ}$ & $-15.46^{\circ}$ & $-15.50^{\circ}$ & $-15.34^{\circ}$ & $-15.46^{\circ}$ & $-15.46^{\circ}$ & $-15.40^{\circ}$ & $-15.34^{\circ}$ \\ 
\hline
\hline
$\Psi^{}_{e1}$ & $-30^{\circ}$ & $-30.00^{\circ}$ & $-29.98^{\circ}$ & $-29.67^{\circ}$ & $-29.62^{\circ}$ & & & & \\ 
\hline
$\Psi^{}_{e2}$ & $-30^{\circ}$ & $-30.00^{\circ}$ & $-30.01^{\circ}$ & $-30.15^{\circ}$ & $-30.18^{\circ}$ & & & & \\ 
\hline
$\Psi^{}_{e3}$ & $60^{\circ}$ & $60.00^{\circ}$ & $59.99^{\circ}$ & $59.82^{\circ}$ & $59.80^{\circ}$ & & & &  \\ 
\hline
$\Psi^{}_{\mu1}$ & $64.86^{\circ}$ & $64.86^{\circ}$ & $64.86^{\circ}$ & $64.76^{\circ}$ & $64.70^{\circ}$ & & & & \\  
\hline
$\Psi^{}_{\mu2}$ & $70.60^{\circ}$ & $70.60^{\circ}$ & $70.61^{\circ}$ & $70.92^{\circ}$ & $70.84^{\circ}$ & & & & \\ 
\hline
$\Psi^{}_{\mu3}$ & $-135.46^{\circ}$ & $-135.46^{\circ}$ & $-135.47^{\circ}$ & $-135.68^{\circ}$ & $-135.54^{\circ}$ & & & & \\ 
\hline
$\Psi^{}_{\tau1}$ & $-124.54^{\circ}$ & $-124.54^{\circ}$ & $-124.55^{\circ}$ & $-124.74^{\circ}$ & $-124.71^{\circ}$ & & & & \\ 
\hline
$\Psi^{}_{\tau2}$ & $-129.90^{\circ}$ & $-129.90^{\circ}$ & $-129.88^{\circ}$ & $-129.56^{\circ}$ & $-129.46^{\circ}$ & & & & \\ 
\hline
$\Psi^{}_{\tau3}$ & $-105.56^{\circ}$ & $-105.56^{\circ}$ & $-105.57^{\circ}$ & $-105.70^{\circ}$ & $-105.83^{\circ}$ & & & & \\ 
\hline
\hline
\end{tabular}
\end{center}
\end{table}

\begin{table}[h]
\caption{Radiative corrections to the Jarlskog ${\cal J}$, the Dirac angle matrix $\Phi$ and the Majorana angle matrix $\Psi$ at $\Lambda \sim 10^{14}$ GeV in the MSSM (with $\tan\beta = 10$) where we choose $\theta^{}_{12} = 34^{\circ}$, $\theta^{}_{23} = 46^{\circ}$, $\theta^{}_{13} = 7^{\circ}$, $\delta = -90^{\circ}$, $\rho = 120^{\circ}$ and $\sigma = 60^{\circ}$ (in the standard parametrization of $V$) as typical inputs at the electroweak energy scale $\Lambda^{}_{\rm EW}$.}
\begin{center}
\begin{tabular}{c|c|c|c|c|c|c|c|c|c}
\hline
\hline
& $\Lambda^{}_{\rm EW} $ & \multicolumn{8}{c}{$\Lambda \sim 10^{14}$ GeV} \\
\cline{3-10}
& & \multicolumn{4}{c|}{Majorana Neutrinos} & \multicolumn{4}{c}{Dirac Neutrinos} \\
\cline{3-10}
& & NH & IH &  \multicolumn{2}{c|}{Near Degeneracy} & NH & IH &  \multicolumn{2}{c}{Near Degeneracy} \\
\cline{5-6} \cline{9-10}
& & $m^{}_{1} \simeq 0$ & $m^{}_{3} \simeq 0$ & $\Delta m^{2}_{31} > 0$ & $\Delta m^{2}_{31} < 0$ & $m^{}_{1} \simeq 0$ & $m^{}_{3} \simeq 0$ & $\Delta m^{2}_{31} > 0$ & $\Delta m^{2}_{31} < 0$ \\
\hline
\hline
${\cal J}$ & -0.027812 & -0.027761 & -0.027682 & -0.020323 & -0.030539 & -0.027786 & -0.027290 & -0.018018 & -0.017959 \\  
\hline
\hline
$\Phi^{}_{e1}$ & $-9.40^{\circ}$ & $-9.38^{\circ}$ & $-9.28^{\circ}$ & $-6.32^{\circ}$ & $-9.68^{\circ}$ & $-9.39^{\circ}$ & $-8.96^{\circ}$ & $-4.71^{\circ}$ & $-4.69^{\circ}$ \\  
\hline
$\Phi^{}_{e2}$ & $-20.50^{\circ}$ & $-20.48^{\circ}$ & $-20.78^{\circ}$ & $-18.15^{\circ}$ & $-27.29^{\circ}$ & $-20.49^{\circ}$ & $-21.52^{\circ}$ & $-36.61^{\circ}$ & $-40.05^{\circ}$ \\
\hline
$\Phi^{}_{e3}$ & $-150.10^{\circ}$ & $-150.14^{\circ}$ & $-149.94^{\circ}$ & $-155.53^{\circ}$ & $-143.03^{\circ}$ & $-150.12^{\circ}$ & $-149.52^{\circ}$ & $-138.68^{\circ}$ & $-135.26^{\circ}$ \\ 
\hline
$\Phi^{}_{\mu1}$ & $-85.46^{\circ}$ & $-85.43^{\circ}$ & $-86.86^{\circ}$ & $-101.86^{\circ}$ & $-100.05^{\circ}$ & $-85.46^{\circ}$ & $-84.96^{\circ}$ & $-75.83^{\circ}$ & $-75.70^{\circ}$ \\
\hline
$\Phi^{}_{\mu2}$ & $-80.10^{\circ}$ & $-80.13^{\circ}$ & $-78.68^{\circ}$ & $-66.29^{\circ}$ & $-63.37^{\circ}$ & $-80.09^{\circ}$ & $-80.32^{\circ}$ & $-82.00^{\circ}$ & $-82.28^{\circ}$ \\
\hline
$\Phi^{}_{\mu3}$ & $-14.44^{\circ}$ & $-14.44^{\circ}$ & $-14.46^{\circ}$ & $-11.85^{\circ}$ & $-16.58^{\circ}$ & $-14.45^{\circ}$ & $-14.72^{\circ}$ & $-22.17^{\circ}$ & $-22.02^{\circ}$ \\
\hline
$\Phi^{}_{\tau1}$ & $-85.14^{\circ}$ & $-85.19^{\circ}$ & $-83.86^{\circ}$ & $-71.82^{\circ}$ & $-70.27^{\circ}$ & $-85.15^{\circ}$ & $-86.08^{\circ}$ & $-99.46^{\circ}$ & $-99.61^{\circ}$ \\ 
\hline
$\Phi^{}_{\tau2}$ & $-79.40^{\circ}$ & $-79.39^{\circ}$ & $-80.54^{\circ}$ & $-95.56^{\circ}$ & $-89.34^{\circ}$ & $-79.42^{\circ}$ & $-78.16^{\circ}$ & $-61.39^{\circ}$ & $-57.67^{\circ}$ \\
\hline
$\Phi^{}_{\tau3}$ & $-15.46^{\circ}$ & $-15.42^{\circ}$ & $-15.60^{\circ}$ & $-12.62^{\circ}$ & $-20.39^{\circ}$ & $-15.43^{\circ}$ & $-15.76^{\circ}$ & $-19.15^{\circ}$ & $-22.72^{\circ}$ \\ 
\hline
\hline
$\Psi^{}_{e1}$ & $-30^{\circ}$ & $-29.97^{\circ}$ & $-30.93^{\circ}$ & $-40.62^{\circ}$ & $-40.72^{\circ}$ & & & & \\ 
\hline
$\Psi^{}_{e2}$ & $-30^{\circ}$ & $-30.01^{\circ}$ & $-29.59^{\circ}$ & $-25.87^{\circ}$ & $-25.56^{\circ}$ & & & & \\ 
\hline
$\Psi^{}_{e3}$ & $60^{\circ}$ & $59.98^{\circ}$ & $60.52^{\circ}$ & $66.49^{\circ}$ & $66.28^{\circ}$ & & & &  \\ 
\hline
$\Psi^{}_{\mu1}$ & $64.86^{\circ}$ & $64.84^{\circ}$ & $65.21^{\circ}$ & $67.56^{\circ}$ & $69.01^{\circ}$ & & & & \\  
\hline
$\Psi^{}_{\mu2}$ & $70.60^{\circ}$ & $70.60^{\circ}$ & $69.87^{\circ}$ & $58.57^{\circ}$ & $65.10^{\circ}$ & & & & \\ 
\hline
$\Psi^{}_{\mu3}$ & $-135.46^{\circ}$ & $-135.44^{\circ}$ & $-135.08^{\circ}$ & $-126.13^{\circ}$ & $-134.11^{\circ}$ & & & & \\ 
\hline
$\Psi^{}_{\tau1}$ & $-124.54^{\circ}$ & $-124.54^{\circ}$ & $-124.07^{\circ}$ & $-118.76^{\circ}$ & $-120.67^{\circ}$ & & & & \\ 
\hline
$\Psi^{}_{\tau2}$ & $-129.90^{\circ}$ & $-129.88^{\circ}$ & $-130.91^{\circ}$ & $-139.58^{\circ}$ & $-142.19^{\circ}$ & & & & \\ 
\hline
$\Psi^{}_{\tau3}$ & $-105.56^{\circ}$ & $-105.58^{\circ}$ & $-105.02^{\circ}$ & $-101.66^{\circ}$ & $-97.14^{\circ}$ & & & & \\ 
\hline
\hline
\end{tabular}
\end{center}
\end{table}

\begin{table}[h]
\caption{Radiative corrections to the Jarlskog ${\cal J}$, the Dirac angle matrix $\Phi$ and the Majorana angle matrix $\Psi$ at $\Lambda \sim 10^{14}$ GeV in the MSSM (with $\tan\beta = 50$) where we choose $\theta^{}_{12} = 34^{\circ}$, $\theta^{}_{23} = 46^{\circ}$, $\theta^{}_{13} = 7^{\circ}$, $\delta = -90^{\circ}$, $\rho = 120^{\circ}$ and $\sigma = 60^{\circ}$ (in the standard parametrization of $V$) as typical inputs at the electroweak energy scale $\Lambda^{}_{\rm EW}$.}
\begin{center}
\begin{tabular}{c|c|c|c|c|c|c|c|c|c}
\hline
\hline
& $\Lambda^{}_{\rm EW} $ & \multicolumn{8}{c}{$\Lambda \sim 10^{14}$ GeV} \\
\cline{3-10}
& & \multicolumn{4}{c|}{Majorana Neutrinos} & \multicolumn{4}{c}{Dirac Neutrinos} \\
\cline{3-10}
& & NH & IH &  \multicolumn{2}{c|}{Near Degeneracy} & NH & IH &  \multicolumn{2}{c}{Near Degeneracy} \\
\cline{5-6} \cline{9-10}
& & $m^{}_{1} \simeq 0$ & $m^{}_{3} \simeq 0$ & $\Delta m^{2}_{31} > 0$ & $\Delta m^{2}_{31} < 0$ & $m^{}_{1} \simeq 0$ & $m^{}_{3} \simeq 0$ & $\Delta m^{2}_{31} > 0$ & $\Delta m^{2}_{31} < 0$ \\
\hline
\hline
${\cal J}$ & -0.027812 & -0.025703 & -0.024026 & -0.013500 & -0.033435 & -0.026677 & -0.010653 & -0.000337 & -0.000332 \\  
\hline
\hline
$\Phi^{}_{e1}$ & $-9.40^{\circ}$ & $-8.52^{\circ}$ & $-7.37^{\circ}$ & $4.55^{\circ}$ & $-95.91^{\circ}$ & $-8.84^{\circ}$ & $-2.61^{\circ}$ & $-0.43^{\circ}$ & $-0.43^{\circ}$ \\  
\hline
$\Phi^{}_{e2}$ & $-20.50^{\circ}$ & $-19.68^{\circ}$ & $-23.88^{\circ}$ & $35.11^{\circ}$ & $-67.78^{\circ}$ & $-20.48^{\circ}$ & $-58.64^{\circ}$ & $-88.07^{\circ}$ & $-101.29^{\circ}$ \\
\hline
$\Phi^{}_{e3}$ & $-150.10^{\circ}$ & $-151.80^{\circ}$ & $-148.75^{\circ}$ & $140.34^{\circ}$ & $-16.31^{\circ}$ & $-150.68^{\circ}$ & $-118.75^{\circ}$ & $-91.50^{\circ}$ & $-78.28^{\circ}$ \\ 
\hline
$\Phi^{}_{\mu1}$ & $-85.46^{\circ}$ & $-84.38^{\circ}$ & $-109.98^{\circ}$ & $16.84^{\circ}$ & $-60.88^{\circ}$ & $-85.24^{\circ}$ & $-61.66^{\circ}$ & $-15.17^{\circ}$ & $-14.67^{\circ}$ \\
\hline
$\Phi^{}_{\mu2}$ & $-80.10^{\circ}$ & $-81.18^{\circ}$ & $-56.79^{\circ}$ & $129.41^{\circ}$ & $-100.27^{\circ}$ & $-79.67^{\circ}$ & $-83.31^{\circ}$ & $-77.09^{\circ}$ & $-78.01^{\circ}$ \\
\hline
$\Phi^{}_{\mu3}$ & $-14.44^{\circ}$ & $-14.44^{\circ}$ & $-13.23^{\circ}$ & $33.75^{\circ}$ & $-18.85^{\circ}$ & $-15.09^{\circ}$ & $-35.03^{\circ}$ & $-87.74^{\circ}$ & $-87.32^{\circ}$ \\
\hline
$\Phi^{}_{\tau1}$ & $-85.14^{\circ}$ & $-87.10^{\circ}$ & $-62.65^{\circ}$ & $158.61^{\circ}$ & $-23.21^{\circ}$ & $-85.92^{\circ}$ & $-115.73^{\circ}$ & $-164.40^{\circ}$ & $-164.90^{\circ}$ \\ 
\hline
$\Phi^{}_{\tau2}$ & $-79.40^{\circ}$ & $-79.14^{\circ}$ & $-99.33^{\circ}$ & $15.48^{\circ}$ & $-11.95^{\circ}$ & $-79.85^{\circ}$ & $-38.05^{\circ}$ & $-14.84^{\circ}$ & $-0.70^{\circ}$ \\
\hline
$\Phi^{}_{\tau3}$ & $-15.46^{\circ}$ & $-13.76^{\circ}$ & $-18.02^{\circ}$ & $5.91^{\circ}$ & $-144.84^{\circ}$ & $-14.23^{\circ}$ & $-26.22^{\circ}$ & $-0.76^{\circ}$ & $-14.40^{\circ}$ \\ 
\hline
\hline
$\Psi^{}_{e1}$ & $-30^{\circ}$ & $-28.75^{\circ}$ & $-45.50^{\circ}$ & $100.87^{\circ}$ & $-28.33^{\circ}$ & & & & \\ 
\hline
$\Psi^{}_{e2}$ & $-30^{\circ}$ & $-29.09^{\circ}$ & $-25.58^{\circ}$ & $158.09^{\circ}$ & $-50.30^{\circ}$ & & & & \\ 
\hline
$\Psi^{}_{e3}$ & $60^{\circ}$ & $57.84^{\circ}$ & $71.08^{\circ}$ & $101.04^{\circ}$ & $78.63^{\circ}$ & & & &  \\ 
\hline
$\Psi^{}_{\mu1}$ & $64.86^{\circ}$ & $64.15^{\circ}$ & $71.85^{\circ}$ & $79.48^{\circ}$ & $128.46^{\circ}$ & & & & \\  
\hline
$\Psi^{}_{\mu2}$ & $70.60^{\circ}$ & $71.77^{\circ}$ & $55.09^{\circ}$ & $-6.43^{\circ}$ & $117.75^{\circ}$ & & & & \\ 
\hline
$\Psi^{}_{\mu3}$ & $-135.46^{\circ}$ & $-135.92^{\circ}$ & $-126.94^{\circ}$ & $-73.05^{\circ}$ & $113.79^{\circ}$ & & & & \\ 
\hline
$\Psi^{}_{\tau1}$ & $-124.54^{\circ}$ & $-124.37^{\circ}$ & $-115.52^{\circ}$ & $-95.97^{\circ}$ & $-147.45^{\circ}$ & & & & \\ 
\hline
$\Psi^{}_{\tau2}$ & $-129.90^{\circ}$ & $-127.91^{\circ}$ & $-148.79^{\circ}$ & $-151.32^{\circ}$ & $-130.03^{\circ}$ & & & & \\ 
\hline
$\Psi^{}_{\tau3}$ & $-105.56^{\circ}$ & $-107.72^{\circ}$ & $-95.69^{\circ}$ & $-112.71^{\circ}$ & $-82.52^{\circ}$ & & & & \\ 
\hline
\hline
\end{tabular}
\end{center}
\end{table}

\begin{table}[h]
\caption{Radiative corrections to the Jarlskog ${\cal J}$, the Dirac angle matrix $\Phi$ and the Majorana angle matrix $\Psi$ at $\Lambda \sim 10^{14}$ GeV in the MSSM (with $\tan\beta = 50$) where we choose $\theta^{}_{12} = 34^{\circ}$, $\theta^{}_{23} = 46^{\circ}$, $\theta^{}_{13} = 7^{\circ}$, $\delta = 90^{\circ}$, $\rho = 120^{\circ}$ and $\sigma = 60^{\circ}$ (in the standard parametrization of $V$) as typical inputs at the electroweak energy scale $\Lambda^{}_{\rm EW}$.}
\begin{center}
\begin{tabular}{c|c|c|c|c|c|c|c|c|c}
\hline
\hline
& $\Lambda^{}_{\rm EW} $ & \multicolumn{8}{c}{$\Lambda \sim 10^{14}$ GeV} \\
\cline{3-10}
& & \multicolumn{4}{c|}{Majorana Neutrinos} & \multicolumn{4}{c}{Dirac Neutrinos} \\
\cline{3-10}
& & NH & IH &  \multicolumn{2}{c|}{Near Degeneracy} & NH & IH &  \multicolumn{2}{c}{Near Degeneracy} \\
\cline{5-6} \cline{9-10}
& & $m^{}_{1} \simeq 0$ & $m^{}_{3} \simeq 0$ & $\Delta m^{2}_{31} > 0$ & $\Delta m^{2}_{31} < 0$ & $m^{}_{1} \simeq 0$ & $m^{}_{3} \simeq 0$ & $\Delta m^{2}_{31} > 0$ & $\Delta m^{2}_{31} < 0$ \\
\hline
\hline
${\cal J}$ & 0.027812 & 0.027568 & 0.015023 & 0.014327 & -0.027134 & 0.026677 & 0.010653 & 0.000337 & 0.000332 \\  
\hline
\hline
$\Phi^{}_{e1}$ & $9.40^{\circ}$ & $9.13^{\circ}$ & $4.46^{\circ}$ & $4.94^{\circ}$ & $-40.79^{\circ}$ & $8.84^{\circ}$ & $2.61^{\circ}$ & $0.43^{\circ}$ & $0.43^{\circ}$ \\  
\hline
$\Phi^{}_{e2}$ & $20.50^{\circ}$ & $21.22^{\circ}$ & $17.53^{\circ}$ & $47.79^{\circ}$ & $-125.26^{\circ}$ & $20.48^{\circ}$ & $58.64^{\circ}$ & $88.07^{\circ}$ & $101.29^{\circ}$ \\
\hline
$\Phi^{}_{e3}$ & $150.10^{\circ}$ & $149.65^{\circ}$ & $158.01^{\circ}$ & $127.27^{\circ}$ & $-13.95^{\circ}$ & $150.68^{\circ}$ & $118.75^{\circ}$ & $91.50^{\circ}$ & $78.28^{\circ}$ \\ 
\hline
$\Phi^{}_{\mu1}$ & $85.46^{\circ}$ & $83.82^{\circ}$ & $35.79^{\circ}$ & $11.03^{\circ}$ & $-124.48^{\circ}$ & $85.24^{\circ}$ & $61.66^{\circ}$ & $15.17^{\circ}$ & $14.67^{\circ}$ \\
\hline
$\Phi^{}_{\mu2}$ & $80.10^{\circ}$ & $80.66^{\circ}$ & $132.66^{\circ}$ & $122.51^{\circ}$ & $-43.25^{\circ}$ & $79.67^{\circ}$ & $83.31^{\circ}$ & $77.09^{\circ}$ & $78.01^{\circ}$ \\
\hline
$\Phi^{}_{\mu3}$ & $14.44^{\circ}$ & $15.52^{\circ}$ & $11.55^{\circ}$ & $46.46^{\circ}$ & $-12.27^{\circ}$ & $15.09^{\circ}$ & $35.03^{\circ}$ & $87.74^{\circ}$ & $87.32^{\circ}$ \\
\hline
$\Phi^{}_{\tau1}$ & $85.14^{\circ}$ & $87.05^{\circ}$ & $139.75^{\circ}$ & $164.03^{\circ}$ & $-14.73^{\circ}$ & $85.92^{\circ}$ & $115.73^{\circ}$ & $164.40^{\circ}$ & $164.90^{\circ}$ \\ 
\hline
$\Phi^{}_{\tau2}$ & $79.40^{\circ}$ & $78.12^{\circ}$ & $29.81^{\circ}$ & $9.70^{\circ}$ & $-11.49^{\circ}$ & $79.85^{\circ}$ & $38.05^{\circ}$ & $14.84^{\circ}$ & $0.70^{\circ}$ \\
\hline
$\Phi^{}_{\tau3}$ & $15.46^{\circ}$ & $14.83^{\circ}$ & $10.44^{\circ}$ & $6.27^{\circ}$ & $-153.78^{\circ}$ & $14.23^{\circ}$ & $26.22^{\circ}$ & $0.76^{\circ}$ & $14.40^{\circ}$ \\ 
\hline
\hline
$\Psi^{}_{e1}$ & $150^{\circ}$ & $148.49^{\circ}$ & $111.98^{\circ}$ & $94.10^{\circ}$ & $-66.18^{\circ}$ & & & & \\ 
\hline
$\Psi^{}_{e2}$ & $150^{\circ}$ & $151.76^{\circ}$ & $161.04^{\circ}$ & $152.04^{\circ}$ & $-20.50^{\circ}$ & & & & \\ 
\hline
$\Psi^{}_{e3}$ & $60^{\circ}$ & $59.75^{\circ}$ & $86.98^{\circ}$ & $113.86^{\circ}$ & $86.68^{\circ}$ & & & &  \\ 
\hline
$\Psi^{}_{\mu1}$ & $55.14^{\circ}$ & $55.54^{\circ}$ & $71.73^{\circ}$ & $78.13^{\circ}$ & $99.09^{\circ}$ & & & & \\  
\hline
$\Psi^{}_{\mu2}$ & $49.40^{\circ}$ & $49.88^{\circ}$ & $10.85^{\circ}$ & $-18.26^{\circ}$ & $148.01^{\circ}$ & & & & \\ 
\hline
$\Psi^{}_{\mu3}$ & $-104.54^{\circ}$ & $-105.42^{\circ}$ & $-82.58^{\circ}$ & $-59.87^{\circ}$ & $112.90^{\circ}$ & & & & \\ 
\hline
$\Psi^{}_{\tau1}$ & $-115.46^{\circ}$ & $-115.33^{\circ}$ & $-103.81^{\circ}$ & $-96.93^{\circ}$ & $-121.70^{\circ}$ & & & & \\ 
\hline
$\Psi^{}_{\tau2}$ & $-110.10^{\circ}$ & $-108.90^{\circ}$ & $-151.62^{\circ}$ & $-150.47^{\circ}$ & $-157.25^{\circ}$ & & & & \\ 
\hline
$\Psi^{}_{\tau3}$ & $-134.44^{\circ}$ & $-135.77^{\circ}$ & $-104.57^{\circ}$ & $-112.60^{\circ}$ & $-81.05^{\circ}$ & & & & \\ 
\hline
\hline
\end{tabular}
\end{center}
\end{table}

\begin{table}[h]
\caption{Radiative corrections to the Jarlskog ${\cal J}$, the Dirac angle matrix $\Phi$ and the Majorana angle matrix $\Psi$ at $\Lambda \sim 10^{14}$ GeV in the MSSM (with $\tan\beta = 50$) where we choose $\theta^{}_{12} = 34^{\circ}$, $\theta^{}_{23} = 46^{\circ}$, $\theta^{}_{13} = 7^{\circ}$, $\delta = 0^{\circ}$, $\rho = 120^{\circ}$ and $\sigma = 60^{\circ}$ (in the standard parametrization of $V$) as typical inputs at the electroweak energy scale $\Lambda^{}_{\rm EW}$.}
\begin{center}
\begin{tabular}{c|c|c|c|c|c|c|c|c|c}
\hline
\hline
& $\Lambda^{}_{\rm EW} $ & \multicolumn{8}{c}{$\Lambda \sim 10^{14}$ GeV} \\
\cline{3-10}
& & \multicolumn{4}{c|}{Majorana Neutrinos} & \multicolumn{4}{c}{Dirac Neutrinos} \\
\cline{3-10}
& & NH & IH &  \multicolumn{2}{c|}{Near Degeneracy} & NH & IH &  \multicolumn{2}{c}{Near Degeneracy} \\
\cline{5-6} \cline{9-10}
& & $m^{}_{1} \simeq 0$ & $m^{}_{3} \simeq 0$ & $\Delta m^{2}_{31} > 0$ & $\Delta m^{2}_{31} < 0$ & $m^{}_{1} \simeq 0$ & $m^{}_{3} \simeq 0$ & $\Delta m^{2}_{31} > 0$ & $\Delta m^{2}_{31} < 0$ \\
\hline
\hline
${\cal J}$ & 0 & 0.001043 & -0.011926 & 0.013144 & -0.036288 & 0 & 0 & 0 & 0 \\  
\hline
\hline
$\Phi^{}_{e1}$ & $0^{\circ}$ & $0.35^{\circ}$ & $-3.67^{\circ}$ & $4.56^{\circ}$ & $-67.38^{\circ}$ & $0^{\circ}$ & $0^{\circ}$ & $0^{\circ}$ & $0^{\circ}$ \\  
\hline
$\Phi^{}_{e2}$ & $0^{\circ}$ & $0.83^{\circ}$ & $-12.41^{\circ}$ & $40.09^{\circ}$ & $-94.47^{\circ}$ & $0^{\circ}$ & $0^{\circ}$ & $0^{\circ}$ & $0^{\circ}$ \\
\hline
$\Phi^{}_{e3}$ & $180^{\circ}$ & $178.82^{\circ}$ & $-163.92^{\circ}$ & $135.35^{\circ}$ & $-18.15^{\circ}$ & $180^{\circ}$ & $180^{\circ}$ & $180^{\circ}$ & $180^{\circ}$ \\ 
\hline
$\Phi^{}_{\mu1}$ & $0^{\circ}$ & $2.03^{\circ}$ & $-26.13^{\circ}$ & $12.83^{\circ}$ & $-88.40^{\circ}$ & $0^{\circ}$ & $0^{\circ}$ & $0^{\circ}$ & $0^{\circ}$ \\
\hline
$\Phi^{}_{\mu2}$ & $180^{\circ}$ & $177.31^{\circ}$ & $-145.45^{\circ}$ & $128.01^{\circ}$ & $-72.50^{\circ}$ & $180^{\circ}$ & $180^{\circ}$ & $180^{\circ}$ & $180^{\circ}$ \\
\hline
$\Phi^{}_{\mu3}$ & $0^{\circ}$ & $0.66^{\circ}$ & $-8.42^{\circ}$ & $39.16^{\circ}$ & $-19.10^{\circ}$ & $0^{\circ}$ & $0^{\circ}$ & $0^{\circ}$ & $0^{\circ}$ \\
\hline
$\Phi^{}_{\tau1}$ & $180^{\circ}$ & $177.62^{\circ}$ & $-150.20^{\circ}$ & $162.61^{\circ}$ & $-24.22^{\circ}$ & $180^{\circ}$ & $180^{\circ}$ & $180^{\circ}$ & $180^{\circ}$ \\ 
\hline
$\Phi^{}_{\tau2}$ & $0^{\circ}$ & $1.86^{\circ}$ & $-22.14^{\circ}$ & $11.90^{\circ}$ & $-13.03^{\circ}$ & $0^{\circ}$ & $0^{\circ}$ & $0^{\circ}$ & $0^{\circ}$ \\
\hline
$\Phi^{}_{\tau3}$ & $0^{\circ}$ & $0.52^{\circ}$ & $-7.66^{\circ}$ & $5.49^{\circ}$ & $-142.75^{\circ}$ & $0^{\circ}$ & $0^{\circ}$ & $0^{\circ}$ & $0^{\circ}$ \\ 
\hline
\hline
$\Psi^{}_{e1}$ & $60^{\circ}$ & $62.12^{\circ}$ & $42.14^{\circ}$ & $93.61^{\circ}$ & $-35.78^{\circ}$ & & & & \\ 
\hline
$\Psi^{}_{e2}$ & $-120^{\circ}$ & $-121.42^{\circ}$ & $-115.33^{\circ}$ & $-12.72^{\circ}$ & $-44.94^{\circ}$ & & & & \\ 
\hline
$\Psi^{}_{e3}$ & $60^{\circ}$ & $59.30^{\circ}$ & $73.19^{\circ}$ & $-80.89^{\circ}$ & $80.72^{\circ}$ & & & &  \\ 
\hline
$\Psi^{}_{\mu1}$ & $60^{\circ}$ & $59.74^{\circ}$ & $71.94^{\circ}$ & $76.22^{\circ}$ & $120.00^{\circ}$ & & & & \\  
\hline
$\Psi^{}_{\mu2}$ & $60^{\circ}$ & $60.44^{\circ}$ & $42.53^{\circ}$ & $179.18^{\circ}$ & $122.03^{\circ}$ & & & & \\ 
\hline
$\Psi^{}_{\mu3}$ & $-120^{\circ}$ & $-120.18^{\circ}$ & $-114.47^{\circ}$ & $104.60^{\circ}$ & $117.97^{\circ}$ & & & & \\ 
\hline
$\Psi^{}_{\tau1}$ & $-120^{\circ}$ & $-119.91^{\circ}$ & $-111.73^{\circ}$ & $-99.22^{\circ}$ & $-127.38^{\circ}$ & & & & \\ 
\hline
$\Psi^{}_{\tau2}$ & $-120^{\circ}$ & $-118.73^{\circ}$ & $-149.88^{\circ}$ & $39.27^{\circ}$ & $-152.44^{\circ}$ & & & & \\ 
\hline
$\Psi^{}_{\tau3}$ & $-120^{\circ}$ & $-121.36^{\circ}$ & $-98.39^{\circ}$ & $59.95^{\circ}$ & $-80.18^{\circ}$ & & & & \\ 
\hline
\hline
\end{tabular}
\end{center}
\end{table}

\begin{figure}
\begin{center}
\vspace{10cm}
\includegraphics[bbllx=6.5cm, bblly=6.0cm, bburx=15.0cm, bbury=14.2cm, width=7cm, height=7cm, angle=0, clip=0]{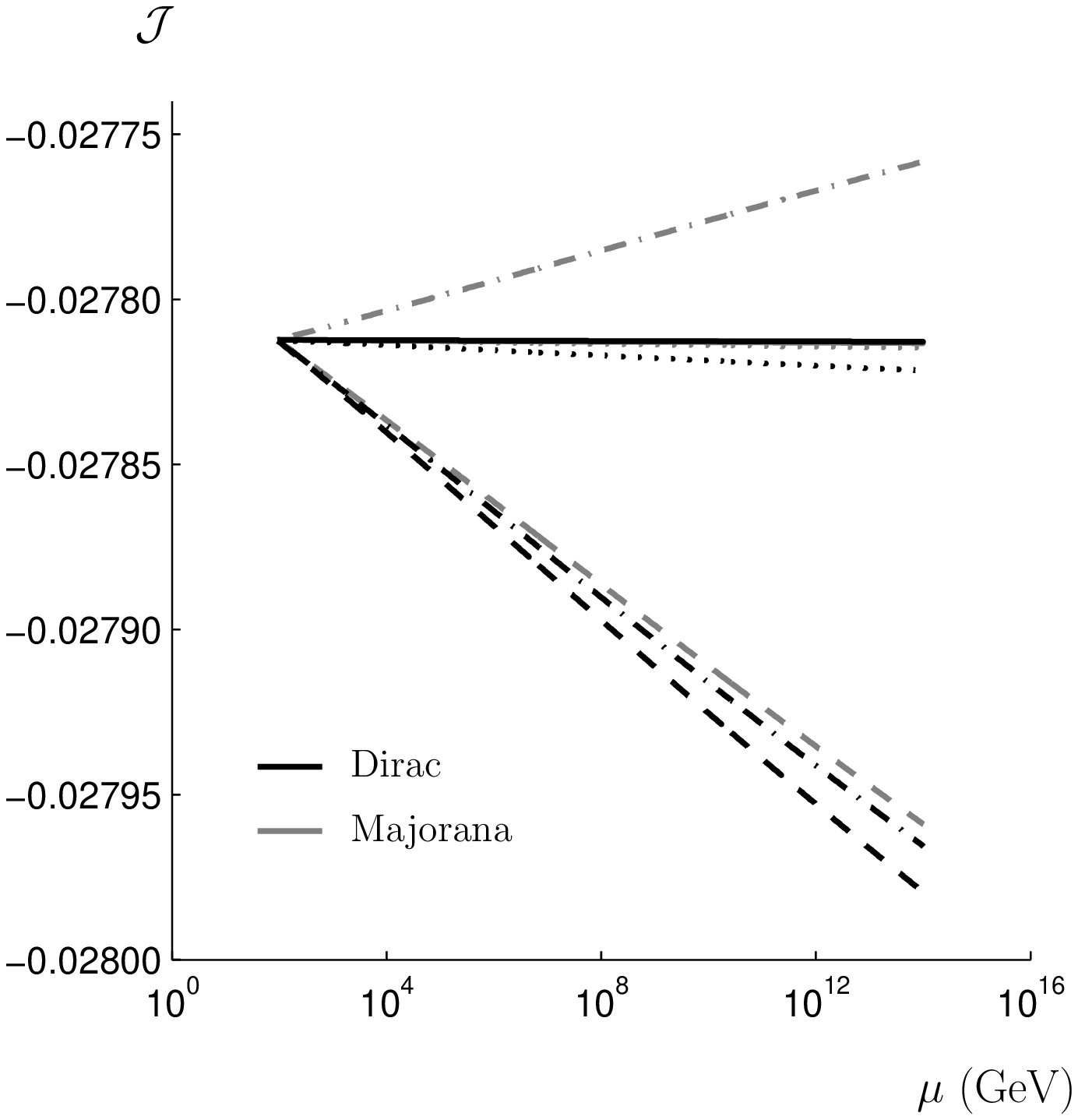}
\vspace{-1.2cm}\caption{Running behaviours of the Jarlskog invariant $\cal J$ from $\Lambda^{}_{\rm EW}$ to $\Lambda \sim 10^{14}$ GeV in the SM for both the Dirac and the Majorana neutrinos, where the corresponding inputs at $\Lambda^{}_{\rm EW}$ and outputs at $\Lambda$ can be found in TABLE I. In this figure, the solid lines stand for the case of NH with $m^{}_{1} \approx 0$, the dotted lines for the case of IH with $m^{}_{3} \approx 0$, the dashed lines for the ND case with $\Delta m^{}_{31} > 0$ and the dash dotted lines for the ND case with $\Delta m^{}_{31} < 0$. Two solid lines in this figure are almost coincide.}
\end{center}
\end{figure}

\begin{figure}
\begin{center}
\vspace{10cm}
\includegraphics[bbllx=6.5cm, bblly=6.0cm, bburx=15.0cm, bbury=14.2cm, width=7cm, height=7cm, angle=0, clip=0]{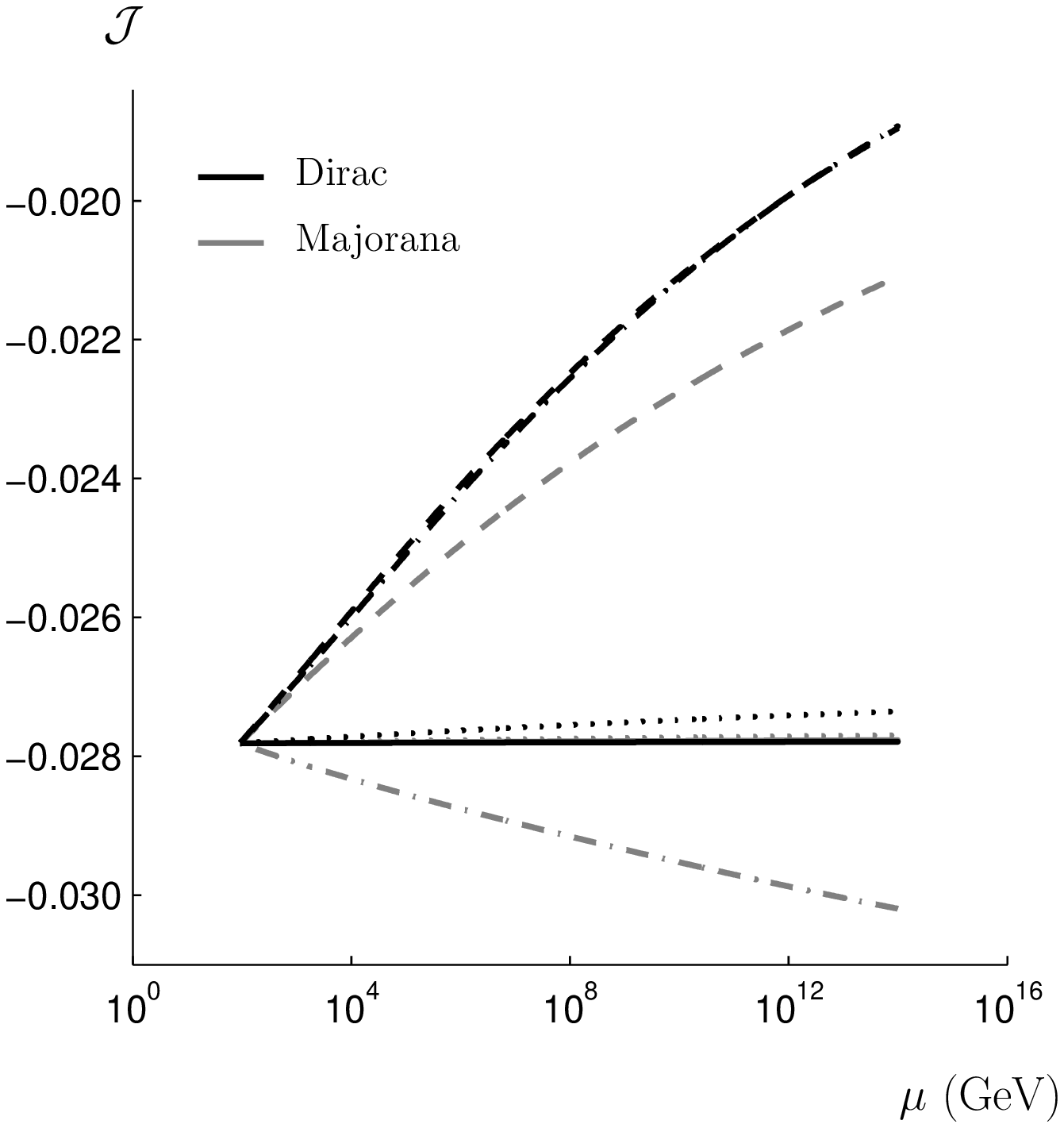}
\vspace{-1.2cm}\caption{Running behaviours of the Jarlskog invariant $\cal J$ from $\Lambda^{}_{\rm EW}$ to $\Lambda \sim 10^{14}$ GeV in the MSSM (with $\tan\beta = 10$) for both the Dirac and the Majorana neutrinos, where the corresponding inputs at $\Lambda^{}_{\rm EW}$ and outputs at $\Lambda$ can be found in TABLE II. In this figure, the solid lines stand for the case of NH with $m^{}_{1} \approx 0$, the dotted lines for the case of IH with $m^{}_{3} \approx 0$, the dashed lines for the ND case with $\Delta m^{}_{31} > 0$ and the dash dotted lines for the ND case with $\Delta m^{}_{31} < 0$. Two solid lines in this figure are almost coincide.}
\end{center}
\end{figure}

\begin{figure}
\begin{center}
\vspace{10cm}
\includegraphics[bbllx=6.5cm, bblly=6.0cm, bburx=15.0cm, bbury=14.2cm, width=7cm, height=7cm, angle=0, clip=0]{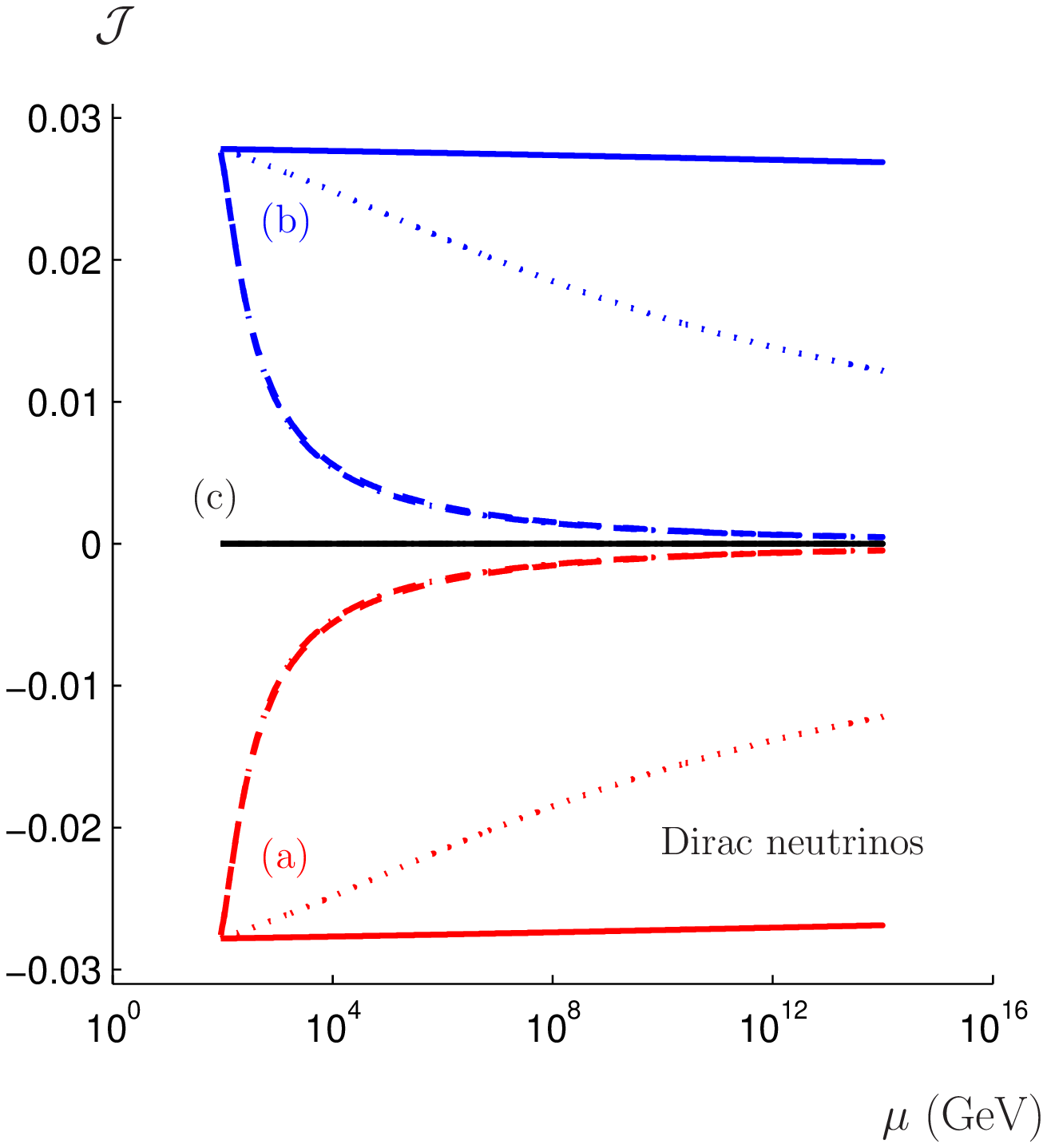}
\vspace{-1.2cm}\caption{Running behaviours of the Jarlskog invariant $\cal J$ from $\Lambda^{}_{\rm EW}$ to $\Lambda \sim 10^{14}$ GeV in the MSSM (with $\tan\beta = 50$) for Dirac neutrinos, where the corresponding inputs at $\Lambda^{}_{\rm EW}$ and outputs at $\Lambda$ for case a), b) and c) can be found in TABLE III, IV and V respectively. The solid lines stand for the case of NH with $m^{}_{1} \approx 0$, the dotted lines for the case of IH with $m^{}_{3} \approx 0$, the dashed lines for the ND case with $\Delta m^{}_{31} > 0$ and the dash dotted lines for the ND case with $\Delta m^{}_{31} < 0$. In this figure, the dashed lines and the dash doted lines are almost coincide in case I and II, and in case III all four lines are almost coincide.}
\end{center}
\end{figure}

\begin{figure}
\begin{center}
\vspace{10cm}
\includegraphics[bbllx=6.5cm, bblly=6.0cm, bburx=15.0cm, bbury=14.2cm, width=7cm, height=7cm, angle=0, clip=0]{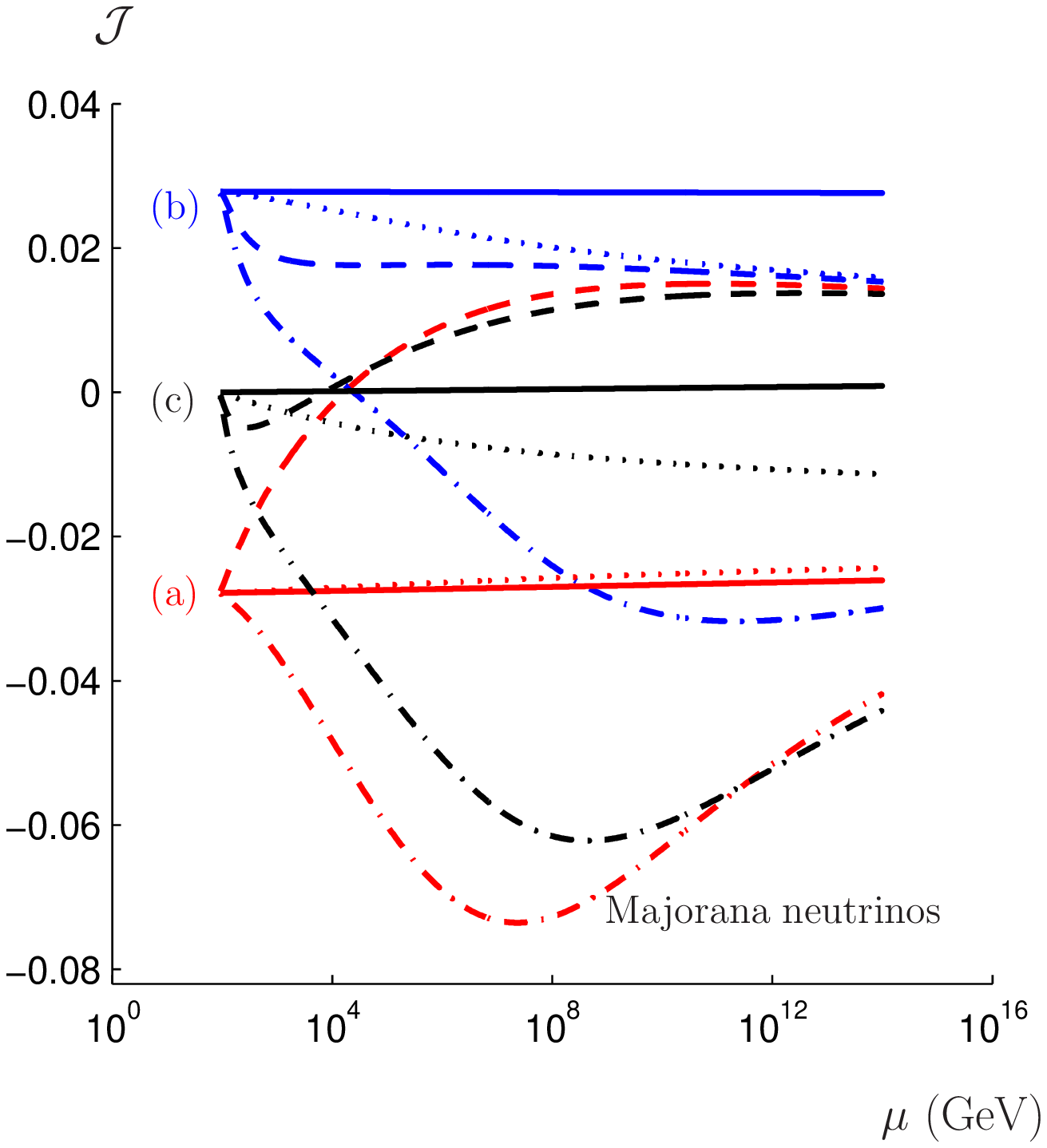}
\vspace{-1.2cm}\caption{Running behaviours of the Jarlskog invariant $\cal J$ from $\Lambda^{}_{\rm EW}$ to $\Lambda \sim 10^{14}$ GeV in the MSSM (with $\tan\beta = 50$) for Majorana neutrinos, where the corresponding inputs at $\Lambda^{}_{\rm EW}$ and outputs at $\Lambda$ for case a), b) and c) can be found in TABLE III, IV and V respectively. In this figure, the solid lines stand for the case of NH with $m^{}_{1} \approx 0$, the dotted lines for the case of IH with $m^{}_{3} \approx 0$, the dashed lines for the ND case with $\Delta m^{}_{31} > 0$ and the dash dotted lines for the ND case with $\Delta m^{}_{31} < 0$.}
\end{center}
\end{figure}

\begin{figure}
\begin{center}
\vspace{10cm}
\includegraphics[bbllx=6.5cm, bblly=6.0cm, bburx=15.0cm, bbury=14.2cm, width=6cm, height=6cm, angle=0, clip=0]{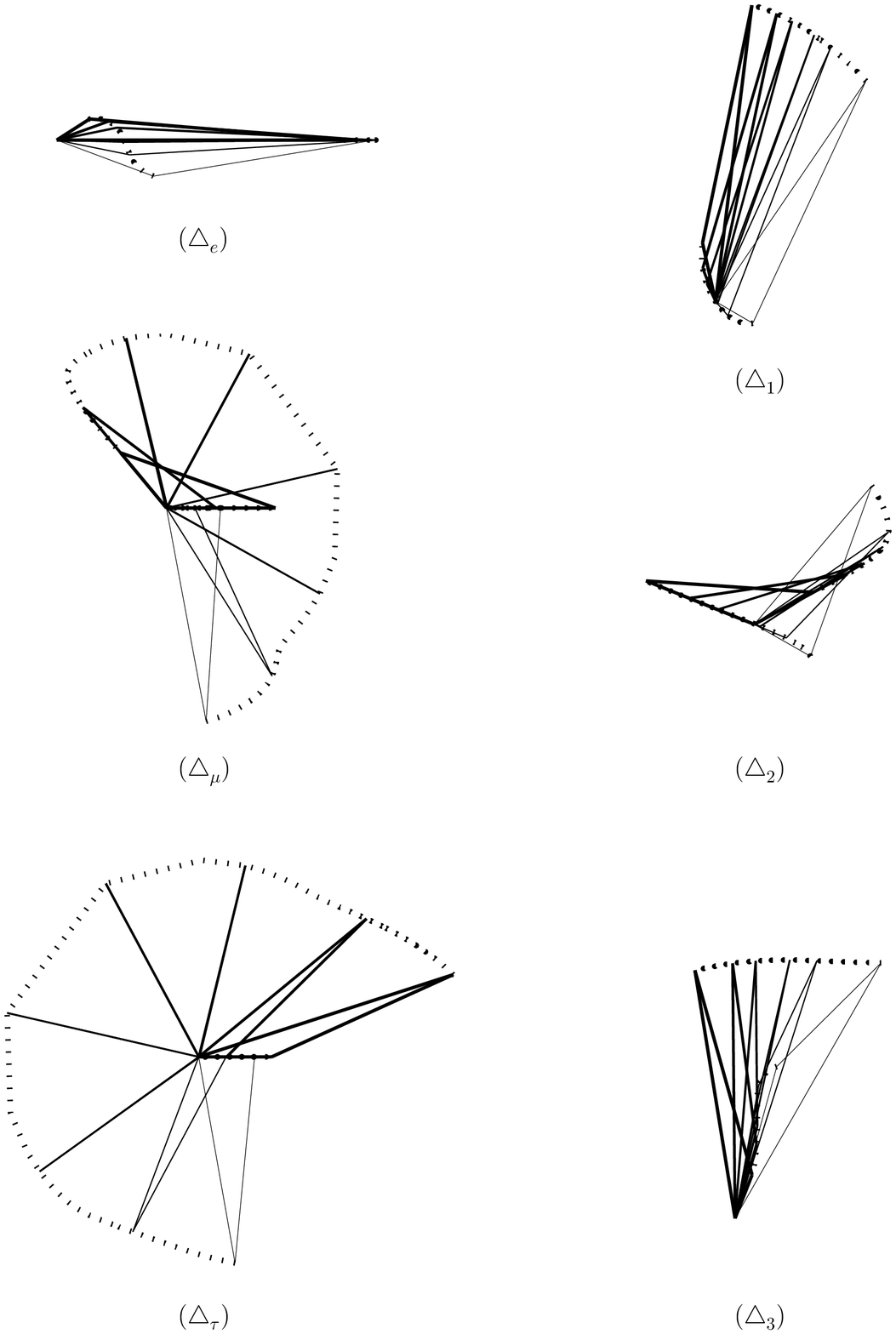}
\vspace{3.6cm}\caption{RG-evolutions of all six leptonic unitarity triangles in the complex plane from $\Lambda^{}_{\rm EW}$ to $\Lambda \sim 10^{14}$ GeV in the MSSM (with $\tan\beta = 50$) for Majorana neutrinos, where triangles with thicker sides are at higher energy scale. Here the neutrino mass spectrum is ND with $\Delta m^{2}_{31} > 0$ and the corresponding inputs at $\Lambda^{}_{\rm EW}$ and outputs at $\Lambda$ can be found in TABLE III. } 
\end{center}
\end{figure}

\begin{figure}
\begin{center}
\vspace{10cm}
\includegraphics[bbllx=6.5cm, bblly=6.0cm, bburx=15.0cm, bbury=14.2cm, width=6cm, height=6cm, angle=0, clip=0]{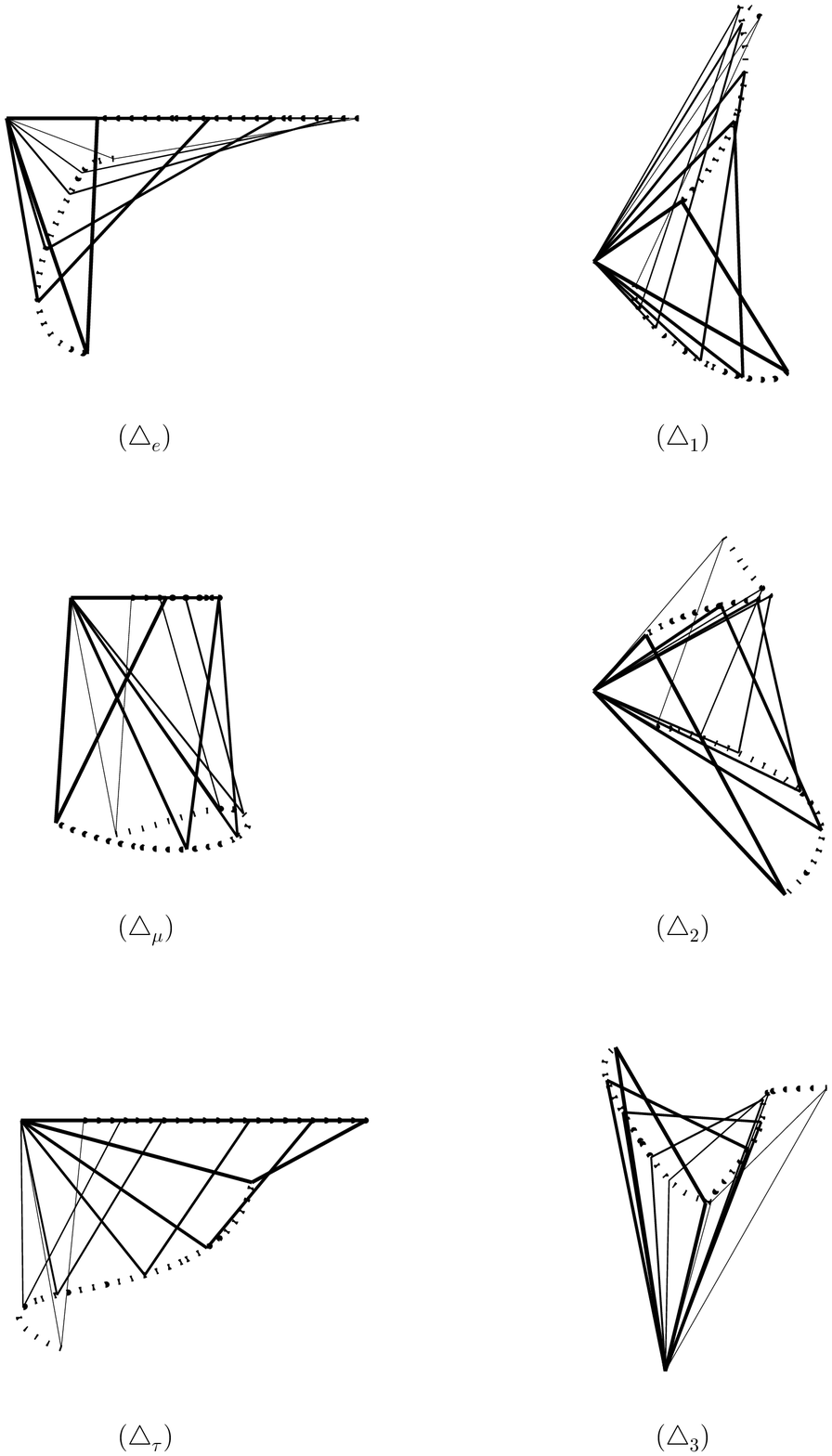}
\vspace{3.6cm}\caption{RG-evolutions of all six leptonic unitarity triangles in the complex plane from $\Lambda^{}_{\rm EW}$ to $\Lambda \sim 10^{14}$ GeV in the MSSM (with $\tan\beta = 50$) for Majorana neutrinos, where triangles with thicker sides are at higher energy scale. Here the neutrino mass spectrum is ND with $\Delta m^{2}_{31} < 0$ and the corresponding inputs at $\Lambda^{}_{\rm EW}$ and outputs at $\Lambda$ can be found in TABLE III. } 
\end{center}
\end{figure}

\end{document}